\documentclass[12pt]{article}
\usepackage{amsfonts,amssymb,pifont,amsthm,dsfont,mathtools,mathrsfs} 

\usepackage[margin=1.25in]{geometry}
\usepackage{setspace}
\usepackage{graphicx}
\usepackage{tikz, tkz-euclide}
\usetikzlibrary{intersections,calc}
\usepackage{caption}
\usepackage{subcaption} 

\usepackage{multirow}
\usepackage{makecell} 
\usepackage{booktabs}
\usepackage{array}
\usepackage{blkarray}

\usepackage{enumerate} 
\usepackage[normalem]{ulem}   
\usepackage{cancel}
\usepackage{verbatim}

\usepackage{appendix}
\usepackage{xcolor}
\usepackage{relsize}
\usepackage{exscale} 

\definecolor{mycolor}{rgb}{0,0,0.75}
\usepackage[round,semicolon]{natbib}
\usepackage[colorlinks, breaklinks,urlcolor=mycolor,citecolor=mycolor,linkcolor=mycolor]{hyperref}

\newtheorem{example}{Example}
\newtheorem{definition}{Definition}
\newtheorem{theorem}{Theorem}

\newtheorem{axiom}{Axiom}

\newtheorem{claim}{Claim}

\newtheorem{proposition}{Proposition}

\begin{document}

\title{Robust Maximum Likelihood Updating\thanks{I am grateful to Yusufcan Masatlioglu, Shaowei Ke, Matias Cattaneo, David Miller, Roc\'{i}o Titiunik, Victor Aguiar, Christopher Chambers, Federico Echenique, Xiaosheng Mu, Collin Raymond, Philipp Strack, Tomasz Strzalecki, and Gerelt Tserenjigmid for their helpful comments.}}

\author{Elchin Suleymanov\thanks{Department of Economics, Mitch Daniels School of Business, Purdue University, West Lafayette, Indiana, USA. Email: \href{mailto:esuleyma@purdue.edu}{esuleyma@purdue.edu}.}
}
\date{\today}

\maketitle

\vspace{-2em}
\begin{abstract}
There is a large body of evidence that decision makers frequently depart from Bayesian updating. This paper introduces a model, \textit{robust maximum likelihood (RML) updating}, where deviations from Bayesian updating are due to multiple priors/ambiguity. Using the decision maker's prior and posteriors as the primitives of the analysis, I axiomatically characterize a representation where the decision maker's probability assessment can be described by a \textit{benchmark prior}, which is interpreted as an initial best guess, and \textit{a set of plausible priors}, which represents all the priors that cannot be ruled out. When new information is received, the decision maker revises her benchmark prior within the set of plausible priors via the \textit{maximum likelihood principle} in a way that ensures maximally \textit{dynamically consistent} behavior, and updates the new benchmark prior using Bayes' rule. I demonstrate how the set of plausible priors can be uniquely identified by comparing ex ante and ex post beliefs and show how most commonly observed updating biases can be accommodated within the model in a unified framework. 

\medskip



\end{abstract}

\thispagestyle{empty}
\clearpage

\setcounter{page}{1}
\pagestyle{plain}

\section{Introduction}
\label{section:intro}

How do decision makers (DMs) update their beliefs when they receive new information? The answer to this question is critical in economic models and policy analyses where one tries to predict the consequences of releasing new information to market participants. The standard assumption in economics is that beliefs are updated using Bayes' rule. However, there is a large body of experimental and empirical evidence showing that decision makers frequently deviate from Bayesian updating. For example, many decision makers tend to underweight base rates (\textit{base rate neglect}), ignore informative signals (\textit{conservatism}), interpret contrary evidence as supportive of their original beliefs (\textit{confirmation bias}), or treat their private information as more informative than it actually is (\textit{overconfidence}).\footnote{For a review of these findings, see, for example, \citet{camerer1995individual}, \citet{rabin1998psychology}, \citet{tversky2004preference}, and \citet{benjamin2019errors}.} This paper introduces a model where deviations from Bayesian updating are due to ambiguity/multiple priors. The model can accommodate the previously mentioned updating biases in a unified framework.

To illustrate how deviations from Bayesian updating can be related to multiple priors, consider the thought experiment due to \citet{ellsberg1961risk} where a DM is told that an urn contains 30 red balls and 60 blue or green balls in an unknown proportion. Let $f_R$, $f_B$, and $f_G$ stand for bets which yield \$100 if the ball drawn from the urn is red, blue, and green, respectively, and \$0 otherwise. When no further information is given, many decision makers are indifferent between these bets, which is consistent with the prior that assigns equal probability to all the colors.\footnote{Ellsberg argued that most decision makers would prefer to bet on red rather than blue or green. While Ellsberg style preferences are common, many experimental findings show that a significant number of decision makers are ambiguity neutral \citep*[see][]{binmore2012much,charness2013ambiguity,stahl2014heterogeneity}. The baseline model in this paper assumes ambiguity neutrality; however, it can be extended to accommodate ambiguity aversion (see the Conclusion).}

Now suppose the experimenter draws a ball from the urn and informs the DM that the ball is not green. How should the DM update her preferences given this information? In particular, should she still be indifferent between $f_R$ and $f_B$? There are two arguments that can be made. First, following the principle of \textit{dynamic consistency}, one can argue that since both $f_R$ and $f_B$ agree on the payoff assigned to the unrealized event (green), the information that this event is ruled out should not affect the original preference. Hence, indifference should be maintained ex post. On the other hand, the information that the ball is not green may suggest that the number of blue balls in the urn is greater than the number of green balls. Since there are only 30 red balls in the urn and 60 blue or green balls, one can also argue that $f_B$ should be preferred to $f_R$ ex post. This preference is incompatible with dynamic consistency, which is the key implication of Bayesian updating. 

The intuition that decision makers may not perform Bayesian updating when they face ambiguity has also been confirmed by experimental findings. For example, in a similar dynamic Ellsberg experiment, \citet*{dominiak2012dynamic} find that a significant number of decision makers whose behavior can be characterized as ambiguity neutral are not Bayesian. Similarly, \citet*{bleichrodt2021testing} find that approximately one in five ambiguity neutral agents violates dynamic consistency. On the other hand, since most existing models tie the DM's ambiguity attitude (rather than ambiguity) to her response to new information, this forces an ambiguity neutral DM to update her beliefs using Bayes' rule. This is inconsistent with the intuition and experimental findings described above, and it is also unnatural, as ambiguity attitude and belief updating are distinct concepts. The model proposed in this paper allows the DM to depart from Bayesian updating when she faces ambiguity, even if her attitude towards ambiguity is neutral. 

While most existing evidence for the updating biases described earlier comes in the context of risk, where agents are given all the necessary information to form a precise prior and there is no objective room for ambiguity, the intuition for why multiple priors might still be relevant is as follows. Even if agents are given sufficient information to form a precise prior, they may still find it difficult to do so either due to computational complexity \citep[][]{zhao2022pseudo} or difficulty in interpreting the given information \citep[][]{epstein2024hard}. Agents who find it difficult to form a precise prior can still use the observed information to form a best guess while bounding the probabilities across states. Agents who simplify the observed information this way realize that they did not use all the relevant information given to them, which naturally leads to multiple priors surrounding their best guess.\footnote{\citet{wilson2014bounded} also models agents who deviate from Bayesian updating due to biases in information processing. In her model, the agent, constrained by her memory capacity, summarizes the observed information using a finite-state automaton.}

To illustrate the model, which is described in Section \ref{section:RML}, let $\Omega$ be a finite set of states, and denote by $\Delta(\Omega)$ the set of all probability measures on $\Omega$. An event $A$ is a nonempty subset of $\Omega$. In the model, the DM is endowed with \textit{a benchmark prior} $\pi\in \Delta(\Omega)$, which is interpreted as the DM's initial best guess, and \textit{a set of plausible priors} $\mathbb{N}$, which represents all the priors the DM cannot rule out. For example, in the Ellsberg experiment, the agent may view the prior that assigns equal probability to all the colors as the benchmark prior, while any prior that assigns 1/3 probability to red can be considered plausible. The benchmark prior $\pi$ represents the best estimate among plausible priors when the agent receives no additional information. 

The updating rule, called \textit{robust maximum likelihood (RML) updating}, can be described by two stages. In the first stage, the DM performs maximum likelihood updating within the set of plausible priors. That is, when the DM learns that event $A$ is realized, she restricts her attention to the subset of plausible priors that maximizes the likelihood of this event. Let $\mathbb{N}_A$ denote this set. Next, the DM chooses a new benchmark prior from this set and updates it using Bayes' rule. If $\mathbb{N}_A$ is a singleton, then the DM chooses the unique likelihood maximizing prior as her new benchmark. If there are multiple priors that maximize the likelihood of the observed event, the DM chooses her new benchmark prior in a way that induces maximally dynamically consistent behavior. Maximal dynamic consistency ensures that the DM stays as close to her original benchmark prior as possible, where closeness is defined in terms of Kullback-Leibler (KL) divergence. When there is no ambiguity (i.e., $\mathbb{N}$ is a singleton), RML updating reduces to Bayesian updating.

To illustrate how RML updating can accommodate a strict preference for $f_B$ over $f_R$ after the realization that the ball drawn from the Ellsberg urn is not green, let $\pi,\pi',\pi''\in \mathbb{N}$, where $\pi$ is the benchmark prior, and $\pi'$ and $\pi''$ represent two plausible priors when there are no green and blue balls in the urn, respectively. When the DM learns that the ball drawn from the urn is not green, maximum likelihood updating implies $\mathbb{N}_A=\{\pi'\}$, where $A=\{R,B\}$. The DM endowed with $\pi'$ as her posterior prefers $f_B$ over $f_R$ despite the fact that she was indifferent between them ex ante.

The rest of this paper proceeds as follows. Section \ref{section:RML} introduces the RML updating rule. In Section \ref{section:representation}, I take the DM's prior and posteriors as the primitives of the analysis and provide a set of axioms that characterizes the updating rule. Section \ref{section:examples} illustrates how the model can explain many well-known updating biases. Section \ref{section:lit_review} provides a discussion on related literature. Section \ref{section:conclusion} concludes. The Appendix includes all the proofs omitted from the main text.

\section{Updating Rule}
\label{section:RML}

Let $\Omega$ be a finite set of states, and denote by $\Delta(\Omega)$ the set of all probability measures on $\Omega$. The collection of all nonempty subsets of $\Omega$ (i.e., events) is denoted by $\mathcal{A}$. The decision maker's \textit{probability assessment} is characterized by $(\pi,\mathcal{P})$ where $\pi\in \Delta(\Omega)$ is her benchmark prior, which is assumed to have full support, and $\mathcal{P}$ is a partition of $\Omega$ that represents the collection of minimal unambiguous events. That is, for any $P\in \mathcal{P}$, the DM assesses that its likelihood is given by $\pi(P)$, and for any nonempty $D\subsetneq P$, the likelihood assigned by $\pi$ reflects the DM's initial best guess. Since $\pi$ is a probability measure, any arbitrary union of the events in $\mathcal{P}$ is also unambiguous.

Since the events in $\mathcal{P}$ are unambiguous, any prior that the DM considers plausible must agree with $\pi$ on $\mathcal{P}$. Hence, it must belong to the set $\mathbb{N}_{\mathcal{P}}(\pi)$ defined by
$$\mathbb{N}_{\mathcal{P}}(\pi)=\{\pi'\in \Delta(\Omega)\mid \pi'(P)=\pi(P) \text{ for all } P\in \mathcal{P}\}.$$
On the other hand, not every prior in $\mathbb{N}_{\mathcal{P}}(\pi)$ is necessarily plausible. For example, if the DM has additional information that can be used to bound the probabilities of ambiguous events, then the set of plausible priors will be a subset of $\mathbb{N}_{\mathcal{P}}(\pi)$.

Formally, let $\mathbb{N}\subseteq \Delta(\Omega)$ denote a closed and convex set of plausible priors.\footnote{Closedness and convexity are standard assumptions in the literature. Closedness requires that if a converging sequence of beliefs is plausible, then the limiting belief is also plausible. Convexity requires that if two beliefs are considered plausible, then any mixture of them is also plausible.} Given a probability assessment $(\pi,\mathcal{P})$, we must have that $\pi\in \mathbb{N}\subseteq \mathbb{N}_{\mathcal{P}}(\pi)$. In addition, to ensure that $\mathcal{P}$ can be interpreted as the set of minimal unambiguous events, it must be the case that any proper subset of a minimal unambiguous event is ambiguous, i.e., for any $P\in \mathcal{P}$ and nonempty $D\subsetneq P$, there exists $\pi'\in \mathbb{N}$ such that $\pi'(D)\neq \pi(D)$. If these two conditions hold, we say that $\mathbb{N}$ is consistent with the probability assessment. 

\begin{definition}
	\label{definition:plausible_priors}
	Let a probability assessment $(\pi,\mathcal{P})$ be given. A closed and convex set of plausible priors $\mathbb{N}\subseteq \Delta(\Omega)$ is consistent with $(\pi,\mathcal{P})$ if (i) $\pi\in \mathbb{N}\subseteq \mathbb{N}_{\mathcal{P}}(\pi)$, and (ii) for any $P\in \mathcal{P}$ and nonempty $D\subsetneq P$, there exists $\pi'\in \mathbb{N}$ such that $\pi'(D)\neq \pi(D)$.
\end{definition}

The case when $\mathbb{N}=\{\pi\}$ represents the extreme where no event is ambiguous. At the other extreme, if $\mathbb{N}=\mathbb{N}_{\mathcal{P}}(\pi)$, then every ``small" ambiguous event (i.e., $D\subsetneq P$ for some $P\in \mathcal{P}$) is maximally ambiguous: there are $\pi',\pi''\in \mathbb{N}$ such that $\pi'(D)=0$ and $\pi''(D)=\pi(P)$. 

For the sake of brevity, if $\mathbb{N}$ is consistent with the probability assessment $(\pi,\mathcal{P})$, we say that $\mathbb{N}$ is a consistent set of plausible priors. In the model, when event $A$ is realized, the DM first restricts her attention to the subset of plausible priors that maximizes the likelihood of the observed event; let $\mathbb{N}_A$ denote this set. Then, within this set, the DM selects a new benchmark prior that is as close to the original benchmark prior as possible, where closeness is defined in terms of Kullback-Leibler (KL) divergence, and updates it using Bayes' rule. The following definition formally describes the updating process.

\begin{definition}
	\label{definition:RML}
	An updating rule is called \textbf{robust maximum likelihood (RML)} updating if, given a probability assessment $(\pi,\mathcal{P})$, where $\pi$ has full support, and a consistent set of plausible priors $\mathbb{N}$, for any $A\in \mathcal{A}$, the posterior $\pi_A$ is given by 
	\begin{align*}
		\pi_A=\underset{\pi_A'\in B(\mathbb{N}_A)}{\arg \min} D_{\text{KL}}(\pi(\cdot|A)\: ||\: \pi_A')
	\end{align*}
	where 
	$$D_{\text{KL}}(\pi(\cdot|A)\:||\: \pi_A' )=-\sum_{\omega\in A}\pi(\omega|A)\ln \left(\frac{\pi_A'(\omega)}{\pi(\omega|A)}\right),$$
	$\pi(\cdot|A)$ is the Bayesian posterior of $\pi$ conditional on the event $A$, and $B(\mathbb{N}_A)$ is the set of all Bayesian posteriors of the priors in $\mathbb{N}_A = \{ \pi'\in \mathbb{N}\mid \pi'(A) = \max_{\pi''\in \mathbb{N}} \pi''(A)\}.$
\end{definition}

The RML updating rule reflects the DM's awareness of potential inaccuracy of her benchmark prior on ambiguous events, which necessitates a revision of the benchmark prior when new information is received, and her willingness to stay as ``close" to her benchmark prior as possible. If all events are unambiguous (i.e., $\mathcal{P}$ is the collection of singletons), RML and Bayesian updating coincide. Overall, RML updating generalizes Bayesian updating by allowing the agent to deviate from it when she lacks full confidence in her benchmark prior.

To ensure that the updating rule is well-behaved, the following richness condition on the set of plausible priors is needed. To introduce the condition, for any prior $\pi'\in \Delta(\Omega)$ and event $A\in \mathcal{A}$, let $\pi'|_{A}$ denote the restriction of $\pi'$ to $A$. For example, if $\Omega=\{\omega_1,\omega_2,\omega_3\}$, $\pi'=(1/2,1/3,1/6)$, and $A=\{\omega_1,\omega_2\}$, then $\pi'|_{A} =(1/2,1/3)$.

\begin{definition}
	\label{definition:richness}
	Let a probability assessment $(\pi,\mathcal{P})$ and a consistent set of plausible priors $\mathbb{N}$ be given. The set of plausible priors $\mathbb{N}$ is \textbf{rich} if the following four conditions hold: 
	\begin{enumerate}
		\item (Non-extremeness) For any nonempty $D\subsetneq P$, where $P\in \mathcal{P}$, there exist $\pi',\pi''\in \mathbb{N}$ such that $$\pi'(D)<\pi(D)<\pi''(D).$$ 
		
		\item (Prior-coherence) For any $A\in \mathcal{A}$, there exists $\pi'\in \mathbb{N}_A$ such that
		$$ \frac{\pi'(\omega)}{\pi'(\omega')}=\frac{\pi(\omega)}{\pi(\omega')} \quad \text{for any } \omega,\omega'\in A\cap P \text{ and } P\in \mathcal{P}.$$
		
		\item (Chain-coherence) For any $P\in \mathcal{P}$ and $D_1\subseteq D_2\subseteq \cdots \subseteq D_m\subseteq P$, there exists $\pi'\in \mathbb{N}$ such that
		$$\pi'(D_i) = \max_{\pi''\in \mathbb{N}}\pi''(D_i) \quad \text{for all } i\in \{1,\ldots,m\}.$$
		
		\item ($\mathcal{P}$-Rectangularity) For any $\pi',\pi''\in \mathbb{N}$ and $P \in \mathcal{P}$, there exist $\pi''',\pi'''' \in \mathbb{N}$ such that 
		$$\pi''' |_{P} = \pi' |_{P}, \quad \pi''' |_{P^c}=\pi'' |_{P^c}, \quad \pi'''' |_{P}=\pi'' |_{P}, \quad  \pi'''' |_{P^c}=\pi' |_{P^c}.$$
	\end{enumerate} 
\end{definition}

The first condition, non-extremeness, ensures that the benchmark prior is not extreme within $\mathbb{N}$, i.e., it assigns neither the highest nor the lowest probability to ambiguous events. The second condition, prior-coherence, requires that for any event $A$, the subset of plausible priors that maximizes the likelihood of the observed event contains a prior that agrees with the benchmark prior on the relative likelihoods of any two states within minimal unambiguous events. The third condition, chain-coherence, requires that for any chain $D_1\subseteq D_2\subseteq \cdots \subseteq D_m$ within a minimal unambiguous event $P$, we can find a plausible prior $\pi'\in \mathbb{N}$ that assigns the maximal probability to all $D_i$ in the given chain. This property is adapted from \cite{bednarski1982binary} and \cite{wasserman1990bayes}, who impose it to ensure that the resulting upper probability satisfies standard structural regularity conditions. The fourth condition, $\mathcal{P}$-rectangularity, requires that for any two plausible priors $\pi',\pi''\in \mathbb{N}$ and minimal unambiguous event $P\in \mathcal{P}$, we can construct two other plausible priors such that the first agrees with $\pi'$ on $P$ and with $\pi''$ on $P^c$, while the second agrees with $\pi''$ on $P$ and with $\pi'$ on $P^c$. In the setup of this paper, this is equivalent to requiring that the set of plausible priors is rectangular with respect to the partition $\mathcal{P}$ in the sense of \cite{epstein2003recursive}.\footnote{While \citet{epstein2003recursive} assume rectangularity to ensure dynamically consistent updating, the RML agent will in general violate dynamic consistency for the events not in $\sigma(\mathcal{P}).$} This property is imposed to ensure that the behavior of plausible priors can be ``separated" across minimal unambiguous events. 

As an example, if $\mathbb{N}=\mathbb{N}_{\mathcal{P}}(\pi)$, then the richness condition is automatically satisfied. However, the richness condition is much weaker than requiring maximal ambiguity for all ``small" ambiguous events. 

Under richness, the RML posteriors can be calculated using a simple formula. To this end, for any $\pi'\in \Delta(\Omega)$, let $\pi'(\cdot|B)$ represent the Bayesian posterior of $\pi'$ conditional on the event $B\in \mathcal{A}$, as before. Given the set of plausible priors $\mathbb{N}$, let the upper probability $\overline{\pi}: \mathcal{A}\rightarrow [0,1]$ be defined by 
$$\overline{\pi}(B) = \max_{\pi'\in \mathbb{N}}\pi'(B) \quad\text{for each } B\in \mathcal{A}.$$ 
If the set of plausible priors is rich, the next proposition shows that the RML updating rule can also be written as
\begin{equation}
	\label{equation:two_stage2}
	\pi_A(\omega)  = \frac{\pi(\omega)}{\sum_{\omega'\in A\cap P_{\omega}}\pi(\omega')} \cdot \frac{\overline{\pi}(A\cap P_\omega)}{\sum_{P\in \mathcal{P}: A\cap P\neq \emptyset}\overline{\pi}(A\cap P)},
\end{equation}
where $P_{\omega}$ is the partition element in $\mathcal{P}$ containing $\omega$. 

\begin{proposition}
	\label{proposition:RML}
	Suppose $\{\pi_A\}_{A\in \mathcal{A}}$ is consistent with RML updating, where $(\pi,\mathcal{P})$ is the DM's probability assessment and $\mathbb{N}$ is a consistent set of plausible priors that is rich. Then, for any $A\in \mathcal{A}$, the RML posterior $\pi_A$ of $\pi$ is given by equation \ref{equation:two_stage2}. 
\end{proposition}

Given equation \ref{equation:two_stage2}, we can see that the RML agent distorts probabilities across minimal unambiguous events while keeping the relative likelihoods consistent within them. For example, if $\mathbb{N}=\mathbb{N}_{\mathcal{P}}(\pi)$, then $\overline{\pi}(A\cap P) = \pi(P)$ must hold for all $P\in \mathcal{P}$ with $A\cap P\neq \emptyset$. In general, we have $\overline\pi(A\cap P)\in \left[\pi(A\cap P), \pi(P)\right]$, where the lower bound is strict if $A\cap P\subsetneq P$, due to the non-extremeness property of $\mathbb{N}$. 

To illustrate the model, consider a modified version of Ellsberg's thought experiment where the DM is told that an urn contains 30 red balls and 60 blue or green balls. Suppose, in addition, the DM is told that the urn contains at least 20 blue and 10 green balls. Then, the set of plausible priors is given by 
$$\mathbb{N}=\{\pi'\in \Delta(\Omega)\mid \pi'(R)=1/3,\,\pi'(B)\in [2/9,5/9]\}.$$
Given the additional information on the composition of the urn, the benchmark prior need not be uniform. For example, one possible benchmark prior is $\pi=(1/3,4/9,2/9)$ if the DM uses the given information to conclude that the best estimate is when there are twice as many blue balls as green balls in the urn. Given the benchmark prior $\pi$ and the set of plausible priors $\mathbb{N}$, the RML posteriors are given by
$$\pi_{\{R,B\}}=(3/8,5/8,0), \quad \pi_{\{R,G\}}=(3/7,0,4/7), \quad \pi_{\{B,G\}}=(0,2/3,1/3).$$

\begin{figure}[htbp]
	\centering
	\begin{subfigure}[b]{0.47\textwidth}
		\centering
		\includegraphics[width=\textwidth]{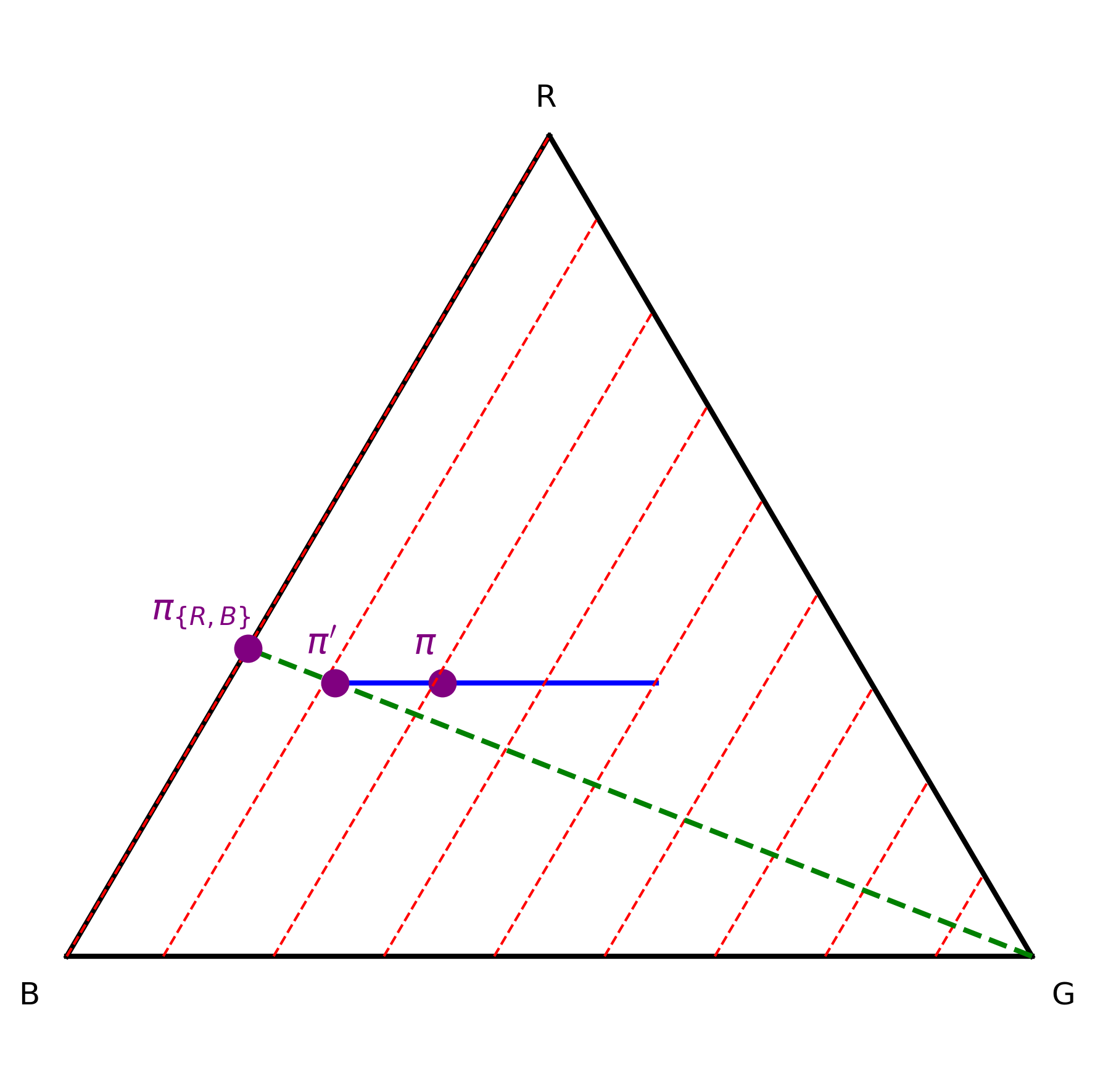}
		\caption{Event $\{R,B\}$ is realized.}
		\label{figure:RB2}
	\end{subfigure}
	\hfill
	\begin{subfigure}[b]{0.47\textwidth}
		\centering
		\includegraphics[width=\textwidth]{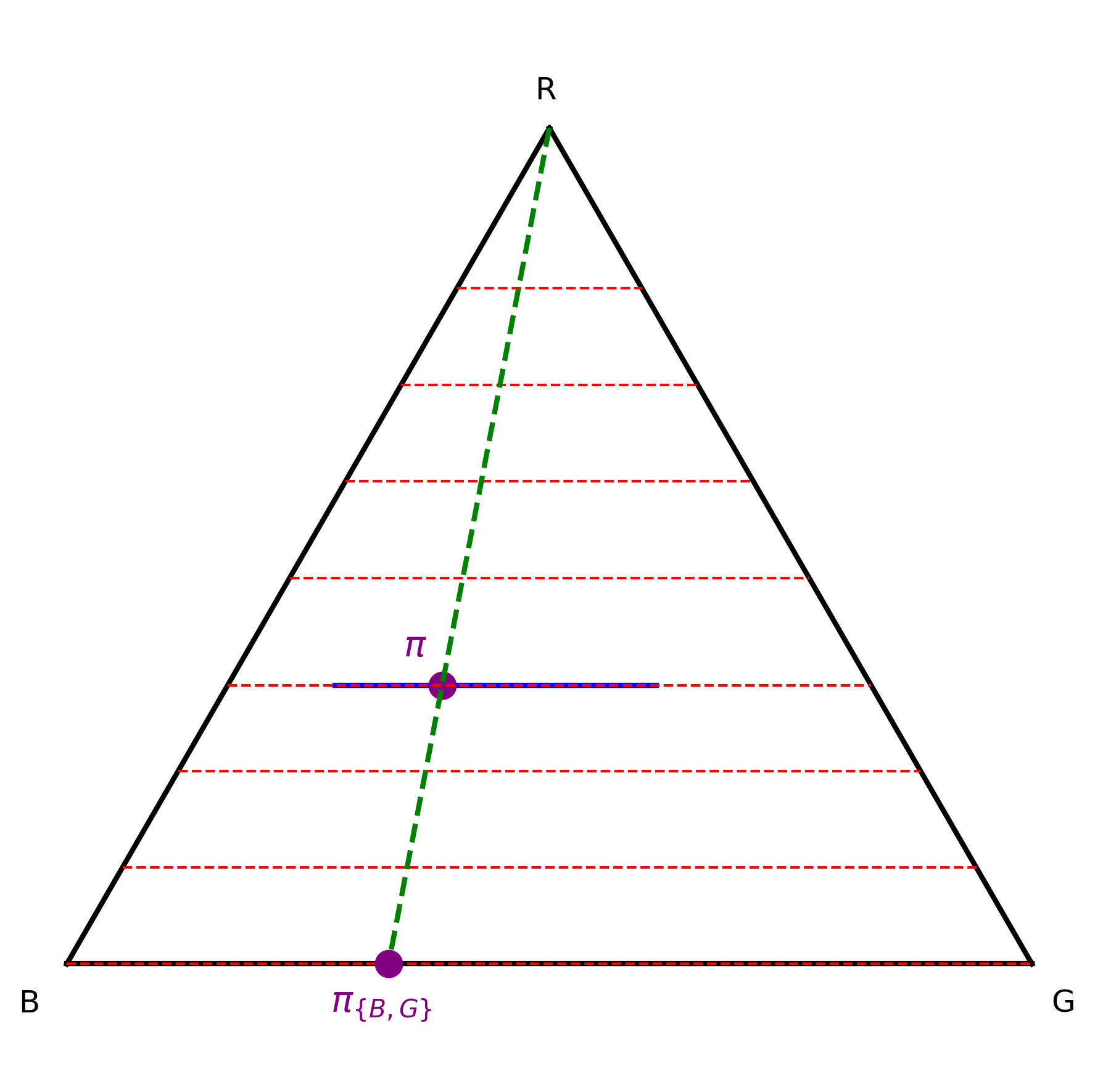}
		\caption{Event $\{B,G\}$ is realized.}
		\label{figure:RG2}
	\end{subfigure}
	\caption{The RML posteriors $\pi_{\{R,B\}}$ and $\pi_{\{B,G\}}$ in the modified Ellsberg example.}
	\label{figure:RML}
\end{figure}

Figure \ref{figure:RML} illustrates the RML posteriors in the modified Ellsberg example for event realizations $\{R,B\}$ and $\{B,G\}$. The blue line in the figure represents the set of plausible priors, and the agent aims to be on the red dashed line that assigns the highest probability to the realized event subject to the constraint imposed by the blue line. When event $\{R,B\}$ is realized, the agent first maximizes the likelihood of $\{R,B\}$, leading her to revise the benchmark prior from $\pi$ to $\pi'$. The agent then performs Bayesian updating using $\pi'$ as her new benchmark prior. When event $\{B,G\}$ is realized, all the points on the blue line correspond to the same red dashed line; hence, the agent performs standard Bayesian updating.

\section{Representation Theorem}
\label{section:representation}

In this section, I provide an axiomatic characterization of RML updating by taking the DM's prior and posteriors $\{\pi_{A}\}_{A\in \mathcal{A}}$ as the primitives of the analysis. These can be interpreted either as directly elicited probability judgments in experimental settings or as beliefs derived from conditional preferences $\{\succcurlyeq_A\}_{A\in \mathcal{A}}$ over acts consistent with the subjective expected utility model \citep{anscombe1963definition}. This section also shows that the set of plausible priors $\mathbb{N}$ and the collection of minimal unambiguous events $\mathcal{P}$ are uniquely identified from the RML agent's updating behavior.

The first axiom is a standard regularity axiom which requires that $\pi_A$ has full support for each $A\in \mathcal{A}$ and that $\pi_A(A)=1$. The assumption that $\pi_A$ has full support is necessary for the updating rule to be well-defined. The second part of the axiom ensures that the agent does not assign positive probability to any state that is already ruled out.

\begin{axiom}[Regularity]	
	\label{axiom:regularity}
	For any $A\in \mathcal{A}$, $\pi_A(\omega)>0$ for all $\omega\in A$ and $\pi_A(A)=1$. 
\end{axiom}

Before introducing the next set of axioms, I first discuss how unambiguous events can be revealed from updating behavior. Suppose an analyst observes that the agent's updating behavior is consistent with Bayes' rule upon realization of some events, while for other events it deviates from Bayesian updating. Intuitively, we would expect updating to be Bayesian for unambiguous events. This is because the realization of such an event is not useful in distinguishing among plausible priors, and hence there is no reason for the DM to deviate from her benchmark prior. As an extreme case, if the agent's updating is always Bayesian, we may conclude that every event is unambiguous for the agent. This argument suggests that we can potentially reveal an unambiguous event from observed posteriors by verifying whether updating is Bayesian when this event is realized. The next example shows that this observation by itself is not sufficient to make such a conclusion. 

\begin{example}
	\label{example:unambiguous}
	Consider a DM who is told that an urn contains 50 red or blue balls and 50 green or yellow balls in unknown proportions. Let $\Omega=\{R,B,G,Y\}$, where $R$, $B$, $G$, and $Y$ stand for states when the ball drawn from the urn is red, blue, green, or yellow, respectively. The set of plausible priors is 
	$$\mathbb{N}=\{\pi'\in \Delta(\{R,B,G,Y\})\mid \pi'(R)+\pi'(B)=\pi'(G)+\pi'(Y)=1/2\}.$$
	In the absence of further information, the DM may choose her benchmark prior as the one that assigns equal probability to all the colors. Now suppose a ball is drawn from the urn and the DM is told that the ball is either blue or green. This information does not favor either blue or green relative to the original information. Hence, we can expect that the benchmark posterior also assigns equal probability to blue and green. But then the agent's updating is Bayesian when $\{B,G\}$ is realized, even though the probability of this event cannot be precisely deduced from the description.
\end{example}

In Example \ref{example:unambiguous}, even though updating is Bayesian when $\{B,G\}$ is realized, it is still possible that $\{B,G\}$ is ambiguous. Now suppose before learning that the ball drawn from the urn is either blue or green, the DM first learns that the ball is not yellow, i.e., the event $\{R,B,G\}$ is realized. The information that the ball is not yellow may suggest that the number of green balls in the urn is greater than the number of yellow balls. Since this information does not say anything regarding the relative proportion of red and blue balls, there is no reason for the DM to deviate from her original evaluation of the relative likelihood of $R$ and $B$. But then the DM's posterior will assign a higher likelihood to $G$ than $B$. This would not be expected if $\{B,G\}$ were unambiguous. 

This example suggests that to verify whether $E$ is an unambiguous event, it is not sufficient that $\pi_E$ is the Bayesian posterior of $\pi$ conditional on the realization of $E$. It must also be the case that $\pi_E$ is the Bayesian posterior of $\pi_A$ for every $A$ that contains  $E$. This motivates the following definition.

\begin{definition}
	\label{definition:ideal}
	An event $E\in \mathcal{A}$ is \textbf{ideal} if it satisfies 
	\begin{equation}
		\label{equation:ideal}
		\pi_E(\omega) = \pi_A(\omega|E) \quad \text{for every } A\in\mathcal A \text{ with } E\subseteq A \text{ and } \omega\in E.
	\end{equation}
	 An event $E\in \mathcal{A}$ is \textbf{symmetrically ideal} if both $E$ and $E^c$ are ideal. By convention, both $\Omega$ and $\emptyset$ are symmetrically ideal. The set $\mathcal{I}$ represents the collection of all symmetrically ideal events.
\end{definition}

The concept of idealness is related to the principle of dynamic consistency that is well-known in the literature. Dynamic consistency requires that if an agent makes plans contingent on the realization of an event $E$, then she should carry out these plans when this event is realized. In the presence of standard axioms, it is established in the literature that dynamic consistency upon realization of an event $E$ is equivalent to the posterior $\pi_E$ being the Bayesian update of $\pi$ (e.g., see \citealp{ghirardato2002revisiting}). Idealness, however, imposes a much stronger version of dynamic consistency, which we can term \textit{perfect dynamic consistency}. Perfect dynamic consistency requires that the agent preserves event $E$-contingent plans not only when $E$ itself is realized but also when any event $A\supset E$ is realized.  Clearly, if the DM is always dynamically consistent so that updating is Bayesian, then the agent will also satisfy perfect dynamic consistency. On the other hand, if the DM's updating is not always Bayesian, we can find events that satisfy dynamic consistency but not perfect dynamic consistency, as shown in Example \ref{example:unambiguous}.

 \cite{gul2014expected} define an event as ideal if the event and its complement satisfy Savage's Sure Thing Principle. Their framework is static, where an analyst observes only ex ante preferences. The notion of symmetric idealness in Definition \ref{definition:ideal} is a natural adaptation of their definition to an environment where we observe both ex ante and ex post preferences.

The next proposition shows that if $\{\pi_A\}_{A\in \mathcal{A}}$ is consistent with RML updating, where the set of plausible priors is rich, then unambiguous events correspond exactly to events that are symmetrically ideal unless $\mathcal{P}= \{\Omega\}$. The case where $\mathcal{P}= \{\Omega\}$ coincides with Bayesian updating, and hence cannot be distinguished from the case where $\mathcal{P}$ is the collection of all singletons (i.e., all events are unambiguous). The reason for requiring symmetric idealness rather than just idealness is due to the fact that the complement of an unambiguous event is also unambiguous, and hence if $E$ is unambiguous, then both $E$ and $E^c$ must be ideal.

\begin{proposition}
	\label{proposition:ideal}
	Suppose $\{\pi_A\}_{A\in \mathcal{A}}$ is consistent with RML updating such that $\mathcal{P}$ is the collection of minimal unambiguous events and the set of plausible priors $\mathbb{N}$ is rich. Then, assuming $\mathcal{P}\neq \{\Omega\}$, an event $E\in \mathcal{A}$ is unambiguous (i.e., $E\in \sigma(\mathcal{P})$) if and only if $E$ is symmetrically ideal. 
\end{proposition}

From the previous proposition, we must have $\mathcal{I} = \sigma(\mathcal{P})$ unless $\mathcal{P}= \{\Omega\}$. Using this result, I will also call the members of $\mathcal{I}$ unambiguous events. The next axiom requires that the collection of symmetrically ideal events forms an algebra.\footnote{A collection of events $\mathcal{I}$ is an algebra over $\Omega$ if (i) $\Omega\in \mathcal{I}$, (ii) $E\in \mathcal{I}$ implies $E^c\in \mathcal{I}$, and (iii) $E,E'\in \mathcal{I}$ implies $E\cap E'\in \mathcal{I}$.} This is necessary for the representation, as unambiguous events are assumed to form an algebra in the model and behaviorally they coincide with symmetrically ideal events.

\begin{axiom}[$\mathcal{I}$-Algebra]
	\label{axiom:algebra2}
	If $E,E'\in \mathcal{I}$, then $E\cap E'\in \mathcal{I}$. 
\end{axiom}

Since $\mathcal{I}$ is an algebra over $\Omega$, there exists a unique partition of the state space that generates it.\footnote{A partition $\mathcal{P}$ of $\Omega$ generates the algebra $\mathcal{I}$ if $E\in \mathcal{I} \Leftrightarrow \text{ there exist } P_1,\dots, P_k\in \mathcal{P}$ such that $P_1\cup \cdots \cup P_k=E$.} Let $\mathcal{P}_{\mathcal{I}}$ denote the partition that generates $\mathcal{I}$. By Proposition \ref{proposition:ideal}, the members of $\mathcal{P}_{\mathcal{I}}$ coincide with the minimal unambiguous events, i.e., any nonempty $D\subsetneq P$ where $P\in \mathcal{P}_{\mathcal{I}}$ is ambiguous. The next four axioms rely on $\mathcal{P}_{\mathcal{I}}$.

To introduce the next axiom, suppose the DM first learns that event $A$ is realized and let $\pi_A$ denote the agent's posterior. Next, suppose the DM learns that a proper subset $D$ of a minimal unambiguous event $P$ is ruled out and let $\pi_{A\setminus D}$ denote the agent's new posterior. The next axiom, betweenness, requires that the agent's posteriors satisfy $\pi_A(P)\geq \pi_{A\setminus D}(P) \geq \pi_A(P|A\setminus D)$. The first inequality reflects monotonicity: an event cannot be considered more likely when there is less evidence supporting it. The second inequality requires that the agent is conservative in revising the probability of unambiguous events: her willingness to bet on $P$ when $D$ is ruled out must be weakly higher than what would be implied by a Bayesian update. This is in line with the intuition that when an ambiguous event $D$ is ruled out, the RML agent revises the benchmark prior to maintain the probability of unambiguous events as close to the original as possible, whereas a Bayesian agent cannot revise the prior ex post.

\begin{axiom}[Betweenness]
	\label{axiom:btw}
	For any $A\in \mathcal{A}$, $P\in \mathcal{P}_{\mathcal{I}}$, and $D\subsetneq A\cap P$, 
	$$\pi_A(P)\geq \pi_{A\setminus D}(P)\geq \pi_A(P|A\setminus D),$$
	where $\pi_A(P|A\setminus D)$ is the Bayesian posterior of $\pi_A$ given the realized event $A\setminus D$. 
\end{axiom}

To introduce the next axiom, for any $A,B\in \mathcal{A}$ with $A\setminus B\neq \emptyset$, let 
$$O_B(A) = \frac{\pi_A(B)}{1-\pi_A(B)}.$$
This reflects the odds for $B$ when event $A$ is realized. Consider a minimal unambiguous event $P\in \mathcal{P}_{\mathcal{I}}$. By the previous axiom, when $D\subsetneq P$ is ruled out, the odds for $P$ decrease by less than they would under a standard Bayesian update. 

Now consider two subsets $D_1$ and $D_2$ of $P$ with an empty intersection. When only $D_1$ is ruled out, consistent with betweenness, the agent might lower the odds for an unambiguous event $P$ minimally and try to keep the probability of $P$ as close to the original value as possible. However, if $D_2$ has already been ruled out, then the effect of removing $D_1$ can be much larger, since keeping the probability of $P$ near its original value becomes more and more difficult as larger portions of $P$ are ruled out. In other words, the odds for a minimal unambiguous event must be increasingly sensitive as the evidence supporting it diminishes. Formally, this implies that the function $O_P(\cdot)$ is submodular on the relevant domain for $P\in \mathcal{P}_{\mathcal{I}}$.

\begin{axiom}[Submodularity]
	\label{axiom:sbm}
	Let $A\in \mathcal{A}$ and $P\in \mathcal{P}_{\mathcal{I}}$ be such that $A\setminus P \neq \emptyset$. Then, for any $D_1, D_2\subseteq P$ with $D_1\cap D_2=\emptyset$,
	$$O_P(A)+O_P(A\setminus (D_1\cup D_2))\leq O_P(A\setminus D_1)+O_P(A\setminus D_2).$$
\end{axiom}

The submodularity axiom addresses how the DM's willingness to bet on a minimal unambiguous event changes as more and more subsets within it are ruled out. The next axiom, log-modularity, addresses situations where the subsets of different partition elements are ruled out. To illustrate, consider two events $A, B\in \mathcal{A}$ and let $O_B(A)$ be defined as before. Let $D_1$ and $D_2$ be such that $D_1\subsetneq A\cap B$ and $D_2\subsetneq A\cap B^c$. Then, Bayesian updating satisfies
$$\ln(O_B(A))+\ln(O_B(A\setminus (D_1\cup D_2)))=\ln(O_B(A\setminus D_1))+\ln(O_B(A\setminus D_2)).$$
That is, log-odds for $B$ are modular when two pieces of contrasting evidence $D_1$ and $D_2$ are removed, where the first favors $B$ while the second is against it. Intuitively, this requires that the informational impacts of removing the evidence for and against $B$ are ``separable" adjustments. While RML updating does not satisfy this property in general, if $B$ is a minimal unambiguous event, then log-modularity must hold for RML updating. 

\begin{axiom}[Log-modularity]
	\label{axiom:lmd}
	Let $A\in \mathcal{A}$ and $P\in \mathcal{P}_{\mathcal{I}}$ be such that $A\cap P\neq \emptyset$ and $A\cap P^c\neq \emptyset$. Then, for any $D_1\subsetneq A\cap P$ and $D_2\subsetneq A\cap P^c$,
	$$\ln(O_P(A))+\ln(O_P(A\setminus (D_1\cup D_2)))=\ln(O_P(A\setminus D_1))+\ln(O_P(A\setminus D_2)).$$
\end{axiom}

The last axiom, idealness, imposes that the agent's updating is consistent with Bayesian updating for all events that are not constrained by the previous axioms. In particular, it requires that every $D\subsetneq P$, where $P\in \mathcal{P}_{\mathcal{I}}$, is ideal.\footnote{Note, however, that $D$ cannot be symmetrically ideal, as it is a proper subset of a minimal unambiguous event.} We can interpret the axiom as requiring that the DM is ``maximally" dynamically consistent by imposing dynamically consistent behavior for events that are not restricted by the previous axioms. 

\begin{axiom}[Idealness]
	\label{axiom:ideal}
	Every proper subset $D\subsetneq P$, where $P\in \mathcal{P}_{\mathcal{I}}$, is ideal.
\end{axiom}

The next theorem provides a characterization result by showing that Axioms \ref{axiom:regularity}-\ref{axiom:ideal} are necessary and sufficient for RML updating. For uniqueness, by Proposition \ref{proposition:ideal}, $\mathcal{P}$ is uniquely revealed as long as $\mathcal{P} \neq \{\Omega\}$. On the other hand, while revealing $\mathcal{P}$ allows us to uniquely reveal $\mathbb{N}_{\mathcal{P}}(\pi)$, this does not immediately reveal the rich set of plausible priors $\mathbb{N}$, since there may be multiple subsets of $\mathbb{N}_{\mathcal{P}}(\pi)$ consistent with the assumptions. Nevertheless, the theorem shows that the rich set of plausible priors $\mathbb{N}$ consistent with the representation is also unique.

\begin{theorem}
	\label{theorem:RML}
	The collection $\{\pi_A\}_{A\in \mathcal{A}}$ satisfies Axioms \ref{axiom:regularity}-\ref{axiom:ideal} if and only if it is consistent with RML updating with a rich set of plausible priors. In addition, the rich set of plausible priors $\mathbb{N}$ consistent with the representation is unique as long as $\mathcal{P} \neq \{\Omega\}$.
\end{theorem}

The next example illustrates the construction for $\mathbb{N}$ in the proof of Theorem \ref{theorem:RML}.

\begin{example}
	\label{example:RML}
	Consider the state space $\Omega = \{\omega_1,\omega_2,\omega_3,\omega_4\}$ with minimal unambiguous events $P_1 = \{\omega_1,\omega_2,\omega_3\}$ and $P_2 = \{\omega_4\}$, which can be revealed from observed updating behavior. Suppose the benchmark prior is $\pi = (\frac{3}{8},\frac{2}{8},\frac{1}{8},\frac{2}{8})$. For any $A\subseteq P_1$, the agent's posterior must be consistent with Bayesian updating. For the remaining non-trivial events, suppose the posteriors are as in the following table. 
	
	{\renewcommand{\arraystretch}{1.5}
		\begin{table}[htbp]
			\centering
			\resizebox{\textwidth}{!}{%
				\begin{tabular}{c|c|c|c|c|c|c}
					\hline
					$A$ & $\{\omega_1,\omega_2,\omega_4\}$ & $\{\omega_1,\omega_3,\omega_4\}$ & $\{\omega_2,\omega_3,\omega_4\}$ & $\{\omega_1,\omega_4\}$ & $\{\omega_2,\omega_4\}$ & $\{\omega_3,\omega_4\}$ \\
					\hline
					$\pi_A$ & $(\frac{9}{20},\frac{6}{20},0,\frac{5}{20})$ & $(\frac{15}{28},0,\frac{5}{28},\frac{8}{28})$ & $(0,\frac{10}{21},\frac{5}{21},\frac{6}{21})$ & $(\frac{2}{3},0,0,\frac{1}{3})$ & $(0,\frac{2}{3},0,\frac{1}{3})$ & $(0,0,\frac{1}{2},\frac{1}{2})$ \\
					\hline
				\end{tabular}
			}
			\caption{Posterior probabilities for events that deviate from Bayesian updating.}
			\label{tab:RMLposteriors}
		\end{table}
	}
	
	We can verify that the observed updating behavior satisfies all the axioms. To proceed with constructing $\mathbb{N}$ from the observed posteriors, we first need to identify the maximal probability $\overline{\pi}(D)$ for each $D\subsetneq P_1$. To illustrate, suppose $D=\{\omega_1,\omega_2\}$. Consider the posterior $\pi_A$ where $A=\{\omega_1,\omega_2,\omega_4\}$. From the representation, we must have that 
	$$\frac{\overline{\pi}(D)}{\pi(P_2)} = \frac{\pi_A(D)}{\pi_A(P_2)} \quad \Rightarrow \quad \overline{\pi}(D) = \frac{\pi_A(D)\pi(P_2)}{\pi_A(P_2)} =\frac{3}{4}.$$
	
	Using this equation, we construct $\overline{\pi}(D)$ for each $D\subsetneq P_1$, as shown in the following table.
	
	{\renewcommand{\arraystretch}{1.5}
		\begin{table}[htbp]
			\centering
			\begin{tabular}{w{c}{1.5cm}|w{c}{1.5cm}|w{c}{1.5cm}|w{c}{1.5cm}|w{c}{1.5cm}|w{c}{1.5cm}|w{c}{1.5cm}}
				\hline
				$D$ & $\{\omega_1,\omega_2\}$ & $\{\omega_1,\omega_3\}$ & $\{\omega_2,\omega_3\}$ & $\{\omega_1\}$ & $\{\omega_2\}$ & $\{\omega_3\}$ \\
				\hline
				$\overline{\pi}(D)$ & $\frac{3}{4}$ & $\frac{5}{8}$ & $\frac{5}{8}$ & $\frac{1}{2}$ & $\frac{1}{2}$ & $\frac{1}{4}$ \\
				\hline
			\end{tabular}
			\caption{Revealed maximal probabilities for each proper subset $D\subsetneq P_1$.}
			\label{tab:RMLmaxpriors}
		\end{table}
	}	
	
	Next, let $\mu:\{1,2,3\} \rightarrow \{1,2,3\}$ denote a permutation of $\{1,2,3\}$, and let $\mathcal{M}$ denote the collection of all such permutations. For example, $\mu = (2,3,1)$ represents the permutation $\mu(1)=2$, $\mu(2)=3$, and $\mu(3)=1$. For each permutation $\mu$, we construct a prior $\pi^{\mu}$ as follows:
	
	\begin{itemize}
		\item First, the state $\omega_{\mu(1)}$ is assigned the highest possible probability, $\overline{\pi}(\omega_{\mu(1)})$, given by the table above.
		\item Next, the state $\omega_{\mu(2)}$ is assigned the probability $\overline{\pi}(\{\omega_{\mu(1)},\omega_{\mu(2)}\}) - \overline{\pi}(\{\omega_{\mu(1)}\})$. That is, $\omega_{\mu(2)}$ gets the residual probability, given the probability assigned to $\omega_{\mu(1)}$ and the maximal possible probability for $\{\omega_{\mu(1)},\omega_{\mu(2)}\}$.
		\item Lastly, $\omega_{\mu(3)}$ is assigned the probability $\overline{\pi}(\{\omega_{\mu(1)},\omega_{\mu(2)},\omega_{\mu(3)}\}) - \overline{\pi}(\{\omega_{\mu(1)},\omega_{\mu(2)}\})$, where $\overline{\pi}(\{\omega_{\mu(1)},\omega_{\mu(2)},\omega_{\mu(3)}\}) = \pi(\{\omega_{\mu(1)},\omega_{\mu(2)},\omega_{\mu(3)}\})$ holds, as $P_1=\{\omega_1,\omega_2,\omega_3\}$ is unambiguous.
	\end{itemize}
	
	The following table illustrates $\pi^{\mu}$ for each $\mu\in \mathcal{M}$.
	
	{\renewcommand{\arraystretch}{1.5}
		\begin{table}[htbp]
			\centering
			\begin{tabular}{c|c|c|c|c|c|c}
				\hline
				$\mu$ & $(1,2,3)$ & $(1,3,2)$ & $(2,1,3)$ & $(2,3,1)$ & $(3,1,2)$ & $(3,2,1)$ \\
				\hline
				$\pi^{\mu}$ & $(\frac{4}{8},\frac{2}{8},0,\frac{2}{8})$ & $(\frac{4}{8},\frac{1}{8},\frac{1}{8},\frac{2}{8})$ & $(\frac{2}{8},\frac{4}{8},0,\frac{2}{8})$ & $(\frac{1}{8},\frac{4}{8},\frac{1}{8},\frac{2}{8})$ & $(\frac{3}{8},\frac{1}{8},\frac{2}{8},\frac{2}{8})$ & $(\frac{1}{8},\frac{3}{8},\frac{2}{8},\frac{2}{8})$ \\
				\hline
			\end{tabular}
			\caption{Priors corresponding to each permutation $\mu\in\mathcal{M}$.}
			\label{tab:RMLpriors}
		\end{table}
	}
	
	The proof of the theorem shows that the collection $\{\pi^{\mu}\}_{\mu\in \mathcal{M}}$ exactly corresponds to the extreme points of the set of plausible priors consistent with the representation. Hence, we can let $\mathbb{N} = \text{co}(\{\pi^{\mu}\}_{\mu\in \mathcal{M}})$, where $\text{co}(\cdot)$ denotes the convex hull operator. By construction, $\mathbb{N}$ satisfies richness and generates the observed posteriors in this example.\footnote{The proof of the theorem uses a slightly different construction at first to derive the representation but then shows that the construction given in this example is valid given the representation.} 
	
	Figure \ref{figure:GRMLcon} illustrates the construction. It describes the posteriors of all the priors in $\mathbb{N}$ conditional on the event realization $P_1$ (i.e., the set $B(\mathbb{N}_{P_1})$, illustrated as the blue shaded area). Since all the plausible priors in $\mathbb{N}$ agree on the likelihood of $\omega_4$, there is a one-to-one mapping between $\mathbb{N}$ and $B(\mathbb{N}_{P_1})$. The extreme points of $B(\mathbb{N}_{P_1})$ correspond to $\pi^{\mu}(\cdot|P_1)$. The intersection of the blue shaded area and the red dashed line describes the Bayesian posteriors of the priors that become relevant when $\omega_2$ is ruled out (e.g., if $\{\omega_1,\omega_3,\omega_4\}$ is realized), and, similarly, the intersection of the blue shaded area and the green dashed line describes the Bayesian posteriors of the priors that become relevant when $\omega_1$ is ruled out (e.g., if $\{\omega_2,\omega_3,\omega_4\}$ is realized). When $\omega_3$ is ruled out, the rightmost side of the blue shaded area represents the Bayesian posteriors of the priors that become relevant in this case.
	
	\begin{figure}[htbp]
		\centering
		\includegraphics[scale=0.50]{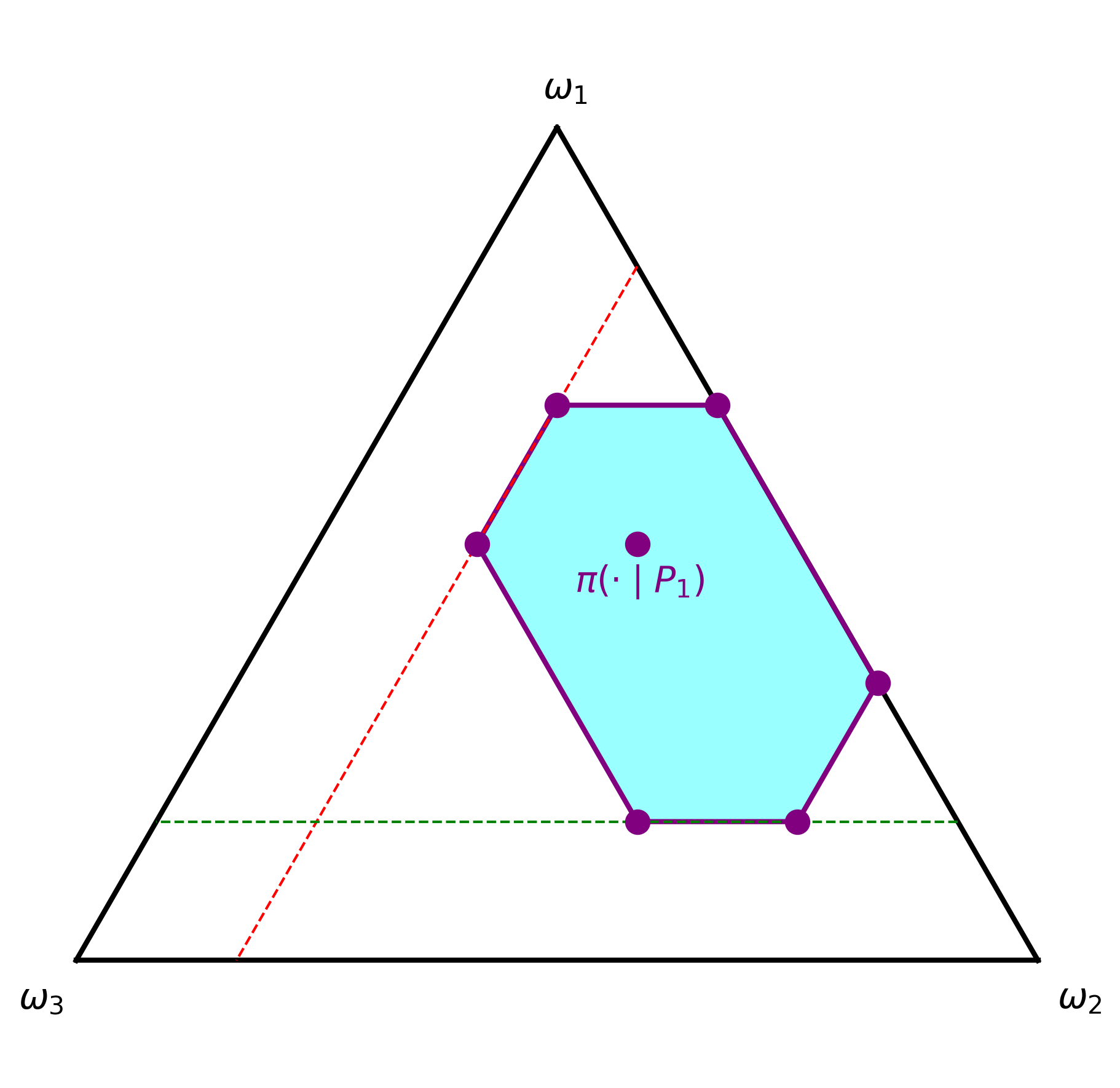}
		\caption{Bayesian posteriors of the priors in $\mathbb{N}$ conditional on the realization of $P_1$.}
		\label{figure:GRMLcon}
	\end{figure}
\end{example}

\section{Applications}
\label{section:examples}

In this section, I show how the RML updating rule can help explain commonly observed biases in belief updating. To introduce the setup, let $\Omega = S\times \Sigma$, where $S=\{s_1,s_2\}$ is the set of payoff-relevant states and $\Sigma=\{\sigma_1,\sigma_2\}$ is the set of possible signal realizations. The DM's benchmark prior $\pi$ is represented by two parameters $(\eta, \alpha)$, where $\eta\geq 1/2$ is the probability that the payoff-relevant state is $s_1$ and $\alpha\geq 1/2$ denotes the probability that the DM receives signal $\sigma_i$ when the payoff-relevant state is $s_i$. The requirement $\eta\geq 1/2$ reflects the DM's initial evaluation that $s_1$ is more likely than $s_2$, and $\alpha\geq 1/2$ indicates that the signal is more likely to match the state.

\begin{figure}[htbp]
	\centering
	\begin{tikzpicture}[scale=1]
		
		\node at (4 , 0.8) {$s_1$};
		\node at (4 , 0) {$s_2$};
		\node at (5.25 , 1.5) {$\sigma_1$};
		\node at (6.5, 1.5) {$\sigma_2$};
		\node at (5.25, 0.8) {$\omega_{11}$};
		\node at (6.5, 0.8) {$\omega_{12}$};
		\node at (5.25, 0) {$\omega_{21}$};
		\node at (6.5, 0) {$\omega_{22}$};

		\draw[color=black] (3.5,1.2)--(7,1.2);
		
		\draw[color=black] (4.4,1.8)--(4.4,-0.3);
		
		\node at (5.75,-1) {\small State Space $\Omega$};
		
		\node at (8.75 , 0.8) {$s_1$};
		\node at (8.75, 0) {$s_2$};
		\node at (10.75 , 1.5) {$\sigma_1$};
		\node at (13.25 , 1.5) {$\sigma_2$};
		\node at (10.75, 0.8) {$\eta\alpha$};
		\node at (13.25, 0.8) {$\eta(1-\alpha)$};
		\node at (10.75, 0) {$(1-\eta)(1-\alpha)$};
		\node at (13.25, 0) {$(1-\eta)\alpha$};

		\draw[color=black] (8.25,1.2)--(14.1,1.2);
		
		\draw[color=black] (9.15,1.8)--(9.15,-0.3);
		
		\node at (11.9,-1) {\small Benchmark Prior $\pi$};
		
	\end{tikzpicture}
\end{figure} 

Given the setup described above, a Bayesian agent who receives signal $\sigma_i$ increases the posterior probability for state $s_i$ and lowers it for $s_j$. We say that an agent over-updates in response to signal $\sigma_i$ if she increases the posterior probability for $s_i$ by more than a Bayesian agent would. Conversely, an agent under-updates if her posterior probability for $s_i$ is lower than that of a Bayesian agent. Table \ref{tab:biases} summarizes the updating biases discussed in this section depending on whether the agent over or under-updates in response to each signal.

\begin{table}[htbp]
	\centering
	\begin{tabular}{@{}lcc@{}} 
		\toprule 
		& \makecell[c]{\textbf{Under-update}\\\textbf{in response to $\sigma_2$}}
		& \makecell[c]{\textbf{Over-update}\\\textbf{in response to $\sigma_2$}} \\
		\midrule 
		\makecell[l]{\textbf{Under-update}\\\textbf{in response to $\sigma_1$}} 
		& Conservatism & Base Rate Neglect \\
		\makecell[l]{\textbf{Over-update}\\\textbf{in response to $\sigma_1$}} 
		& Confirmation Bias & Overconfidence \\
		\bottomrule 
	\end{tabular}
	\caption{Classification of belief updating biases.}
	\label{tab:biases} 
\end{table}

I will next formally introduce each updating bias and show how RML updating can accommodate it. 

\subsection{Confirmation Bias and Conservatism}
\label{section:CCbias}

Let \(\pi(s_i|\sigma_j)\) denote the Bayesian posterior that the state is \(s_i\) when \(\sigma_j\) is realized for \(i,j\in \{1,2\}\). Let \(\pi_{\sigma_1}\) and \(\pi_{\sigma_2}\) denote the observed posteriors. Since the agent originally considers \(s_1\) to be more likely than \(s_2\), we say that the agent exhibits confirmation bias if 
$$\pi_{\sigma_1}(s_1)>  \pi(s_1|\sigma_1) \quad \text{and} \quad   \pi_{\sigma_2}(s_1)> \pi(s_1|\sigma_2).$$
That is, the agent interprets any information as more favorable to state $s_1$ relative to the Bayesian agent.
In addition, since \(\sigma_i\) is more informative of the state \(s_i\), we say that the agent exhibits conservatism if 
$$\pi(s_1)\leq \pi_{\sigma_1}(s_1)< \pi(s_1|\sigma_1) \quad \text{and} \quad  \pi(s_2)\leq \pi_{\sigma_2}(s_2)< \pi(s_2|\sigma_2).$$
In other words, the agent updates in the relevant direction by less than the Bayesian agent.
Notice that \(\pi_{\sigma_2}(s_1)> \pi(s_1|\sigma_2)\) if and only if \(\pi_{\sigma_2}(s_2)< \pi(s_2|\sigma_2)\). Hence, when \(\sigma_2\) is realized, an agent who exhibits conservatism also exhibits confirmation bias. Confirmation bias can be more extreme, however, as it allows for the possibility that \(\pi(s_2)> \pi_{\sigma_2}(s_2)\) (or equivalently, \(\pi_{\sigma_2}(s_1)>\pi(s_1)\)), meaning that the agent treats contrary evidence as even more supportive of the original beliefs. However, when \(\sigma_1\) is realized, agents who exhibit confirmation bias and conservatism move in opposite directions relative to the Bayesian update. Both confirmation bias and conservatism are frequently reported in experiments (see \citealp{lord1979biased, darley1983hypothesis} for classical evidence on confirmation bias, and \citealp{edwards1968conservatism} for classical findings on conservatism).

To accommodate confirmation bias and conservatism within RML updating, consider an agent who is uncertain regarding the signal accuracy. Specifically, the agent considers it plausible that the signal accuracy is potentially distinct from $\alpha$ and depends on the state realization, as illustrated below.

\begin{figure}[ht!]
	\centering
	\begin{tikzpicture}[scale=1]
		\node at (8.75 , 0.8) {$s_1$};
		\node at (8.75, 0) {$s_2$};
		\node at (10.75 , 1.5) {$\sigma_1$};
		\node at (13.25 , 1.5) {$\sigma_2$};
		\node at (10.75, 0.8) {$\eta\beta$};
		\node at (13.25, 0.8) {$\eta(1-\beta)$};
		\node at (10.75, 0) {$(1-\eta)(1-\gamma)$};
		\node at (13.25, 0) {$(1-\eta)\gamma$};
		\draw[color=black] (8.25,1.2)--(14.1,1.2);
		\draw[color=black] (9.15,1.8)--(9.15,-0.3);
		\node at (11.5,-1) {\small Alternative Plausible Prior $\pi^{\beta,\gamma}$};
	\end{tikzpicture}
\end{figure} 

The parameter $\beta\in  [\underline{\beta},\overline{\beta}]$ reflects the signal accuracy in state $s_1$, while $\gamma\in [\underline{\gamma},\overline{\gamma}]$ reflects the signal accuracy in state $s_2$. The set of plausible priors is given by 
$$\mathbb{N}_{\beta,\gamma} = \{\pi^{\beta,\gamma}\mid \beta \in [\underline{\beta},\overline{\beta}], \gamma\in [\underline{\gamma},\overline{\gamma}], \min\{\underline{\beta}, \underline{\gamma}\}\geq 0, \max\{\overline{\beta}, \overline{\gamma}\}\leq 1, \alpha \in [\underline{\beta},\overline{\beta}]\cap [\underline{\gamma},\overline{\gamma}] \}.$$
The requirements $\min\{\underline{\beta}, \underline{\gamma}\}\geq 0$ and $\max\{\overline{\beta}, \overline{\gamma}\}\leq 1$ ensure that all the probabilities are non-negative. Notice that $\pi\in \mathbb{N}_{\beta,\gamma}$, as $\alpha \in [\underline{\beta},\overline{\beta}]\cap [\underline{\gamma},\overline{\gamma}]$.

Given this set of plausible priors, we can calculate the RML posteriors as 
\begin{align*}
	\pi_{\sigma_1}(s_1) &= \frac{\eta\overline{\beta}}{\eta\overline{\beta}+(1-\eta)(1-\underline{\gamma})}, &
	\pi_{\sigma_1}(s_2) &= \frac{(1-\eta)(1-\underline{\gamma})}{\eta\overline{\beta}+(1-\eta)(1-\underline{\gamma})}, \\[8pt]
	\pi_{\sigma_2}(s_1)  &= \frac{\eta(1-\underline{\beta})}{\eta(1-\underline{\beta})+(1-\eta)\overline{\gamma}}, &
	\pi_{\sigma_2}(s_2)  &= \frac{(1-\eta)\overline{\gamma}}{\eta(1-\underline{\beta})+(1-\eta)\overline{\gamma}}.
\end{align*}
On the other hand, the Bayesian posteriors are given by
\begin{align*}
	\pi(s_1|\sigma_1) &= \frac{\eta\alpha}{\eta\alpha+(1-\eta)(1-\alpha)}, &
	\pi(s_2|\sigma_1)  &= \frac{(1-\eta)(1-\alpha)}{\eta\alpha+(1-\eta)(1-\alpha)}, \\[8pt]
	\pi(s_1|\sigma_2)  &= \frac{\eta(1-\alpha)}{\eta(1-\alpha)+(1-\eta)\alpha}, &
	\pi(s_2|\sigma_2)&= \frac{(1-\eta)\alpha}{\eta(1-\alpha)+(1-\eta)\alpha}.
\end{align*}

From these expressions, RML updating results in confirmation bias if
$$\frac{\overline{\gamma}}{1-\underline{\beta}+\overline{\gamma}}< \alpha < \frac{\overline{\beta}}{1+\overline{\beta}-\underline{\gamma}}.$$
Conversely, RML updating results in conservatism if
$$\overline{\beta}\geq 1-\underline{\gamma}, \quad \overline{\gamma}\geq 1-\underline{\beta}, \quad \text{and}\quad \alpha  > \max \left\{\frac{\overline{\beta}}{1+\overline{\beta}-\underline{\gamma}}, \frac{\overline{\gamma}}{1-\underline{\beta}+\overline{\gamma}}\right\}.$$
Notice that the first two inequalities are satisfied if we assume $\min\{\underline{\beta}, \underline{\gamma}\}\geq 1/2$, as $\min\{\overline{\beta}, \overline{\gamma}\}\geq \alpha\geq 1/2$. The requirement $\min\{\underline{\beta}, \underline{\gamma}\}\geq 1/2$ is consistent with the original assumption that $\sigma_i$ is more informative for state $s_i$. This also demonstrates that the DM will exhibit extreme versions of confirmation bias within RML updating (i.e., $\pi_{\sigma_2}(s_1)>\pi(s_1)$) only if she does not rule out the possibility that $\sigma_i$ might be less informative for $s_i$. Furthermore, the higher $\alpha$ is, the more likely it is that the last inequality will be satisfied, resulting in conservatism. This is consistent with the meta-analysis reported in \cite{benjamin2019errors} which finds that higher signal accuracy (diagnosticity) results in more conservatism (underinference) on average. 

\begin{figure}[htbp]
	\centering
	\includegraphics[scale=0.50]{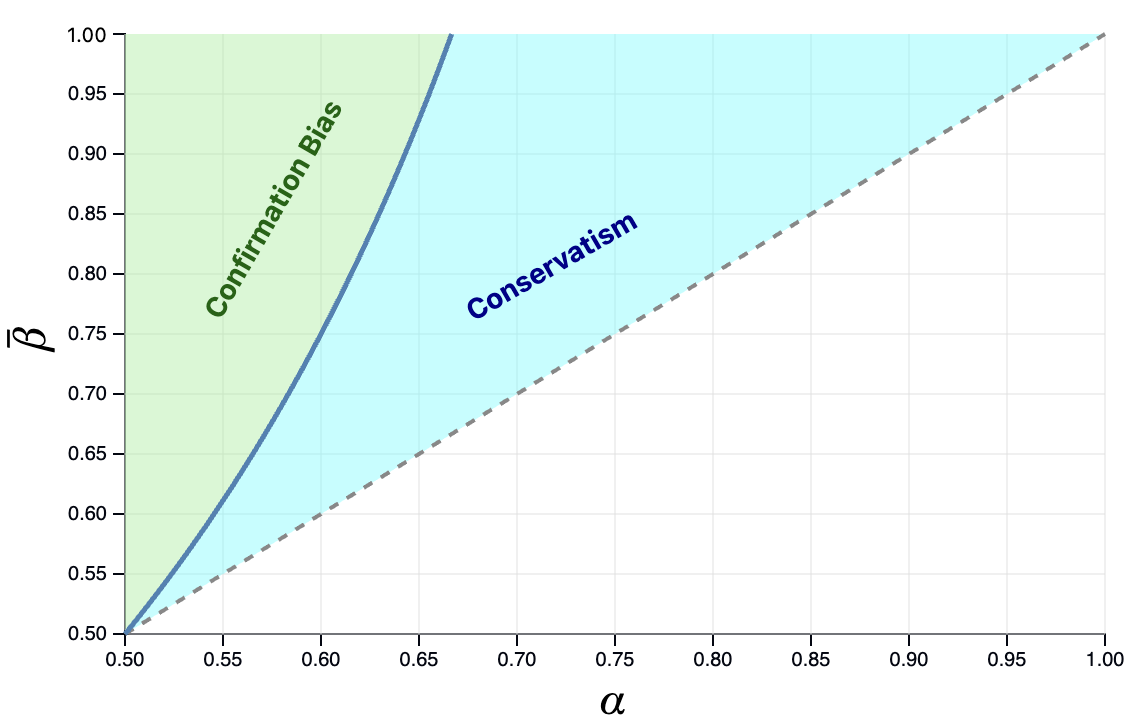}
	\caption{Parameters consistent with confirmation bias and conservatism when $\sigma_1$ is realized, assuming $\underline{\gamma}=1/2$.}
	\label{figure:CCBias}
\end{figure}

To illustrate, suppose the parameters satisfy the first two inequalities for conservatism and set $\underline{\gamma}=1/2$. When $\sigma_1$ is realized, confirmation bias arises if
$$\alpha < \frac{2\overline{\beta}}{1+2\overline{\beta}},$$
while conservatism implies the opposite. Figure \ref{figure:CCBias} illustrates the areas corresponding to confirmation bias and conservatism in this case. Since $\overline{\beta}\leq 1$, the above threshold is at most 2/3. Hence, when $\alpha>2/3$, the agent exhibits conservatism regardless of the value of $\overline{\beta}$. However, when $\alpha<2/3$, the direction of the bias depends on the precise value of $\overline{\beta}$. If the agent considers it plausible that the signal is highly accurate in state $s_1$ (high $\overline{\beta}$), she exhibits confirmation bias. Conversely, if the maximal plausible signal accuracy is low, she exhibits conservatism.

\subsection{Base Rate Neglect and Overconfidence}
\label{section:BOBias}

Base rate neglect refers to the phenomenon where decision makers tend to underweight the base rate, $\eta$, when they form their posteriors. \citet{kahneman1973psychology} and \citet{bar1980base} show this finding in a series of experiments, the most famous of which is known as the ``cab problem.'' In the experiment, DMs are told that there are two cab companies, Blue and Green, one of which has been involved in a hit-and-run accident. The proportion of Blue cabs in the city is 85\%, and the cab involved in the accident was identified as Green by a witness who is accurate 80\% of the time (that is, $\eta=0.85$ and $\alpha=0.8$). When DMs are asked to predict the probability that the car involved in the accident is Green, the median and modal response is 0.8, much higher than the Bayesian posterior ($\approx 0.41$). 

To introduce base rate neglect formally, let $\pi_{1/2}(s_i|\sigma_j)$ denote the Bayesian posterior that the state is $s_i$ upon the realization of $\sigma_j$ assuming that $\eta=1/2$. Notice that $\pi_{1/2}(s_i|\sigma_i)=\alpha$ and $\pi_{1/2}(s_i|\sigma_j)=1-\alpha$ if $i\neq j$. We say that the agent exhibits base rate neglect if, for any $i,j\in \{1,2\}$,
$$\pi_{\sigma_j}(s_i) = \lambda_j \pi_{1/2}(s_i|\sigma_j) + (1-\lambda_j) \pi(s_i|\sigma_j)$$
for some $\lambda_j\in (0,1]$, which can potentially depend on signal realization. The higher $\lambda_j$ is, the larger the extent of base rate neglect. The cab experiment suggests that even $\lambda_j=1$ (complete base rate ignorance) is possible. Since the condition holds for $\pi_{\sigma_i}(s_j)$ whenever it holds for $\pi_{\sigma_i}(s_i)$, using the fact that $\pi_{1/2}(s_i|\sigma_i)=\alpha$ and $\eta\geq 1/2$, we can also express base rate neglect as
$$\alpha \leq \pi_{\sigma_1}(s_1)  < \pi(s_1|\sigma_1) \quad \text{and} \quad \alpha \geq \pi_{\sigma_2}(s_2) >  \pi(s_2|\sigma_2).$$
Hence, the agent under-updates when she receives signal $\sigma_1$ and over-updates when she receives signal $\sigma_2$.

Alternatively, decision makers who treat their private information as more precise than it actually is are described as overconfident \citep[see][for a review of psychology literature on overconfidence and its implications for asset markets]{odean1998volume}. As an illustration, suppose $s_1=\text{good market}$, $s_2=\text{bad market}$, $\sigma_1=\text{good jobs report}$, and $\sigma_2=\text{bad jobs report}$. Overconfident investors tend to over-invest when they observe a good jobs report and under-invest when they observe a bad jobs report. Formally, the decision maker exhibits overconfidence if
$$\pi_{\sigma_1}(s_1)>  \pi(s_1|\sigma_1) \quad \text{and} \quad   \pi_{\sigma_2}(s_2)> \pi(s_2|\sigma_2).$$
Hence, an overconfident agent always over-updates. 

Notice that when $\sigma_2$ is realized, an agent who exhibits base rate neglect also exhibits overconfidence. However, when $\sigma_1$ is realized, base rate neglect results in a posterior that moves in the opposite direction to that of an overconfident agent (relative to the Bayesian benchmark). This parallels the relationship between confirmation bias and conservatism.

To accommodate base rate neglect and overconfidence within RML updating, consider an agent who trusts the signal accuracy but perceives uncertainty regarding the base rate. Specifically, the agent considers it plausible that the base rate is potentially distinct from $\eta$ and depends on whether the signal matches the state, as illustrated below.

\begin{figure}[htbp]
	\centering
	\begin{tikzpicture}[scale=1]
		\node at (8.75 , 0.8) {$s_1$};
		\node at (8.75, 0) {$s_2$};
		\node at (10.75 , 1.5) {$\sigma_1$};
		\node at (13.25 , 1.5) {$\sigma_2$};
		\node at (10.75, 0.8) {$\theta\alpha$};
		\node at (13.25, 0.8) {$\delta(1-\alpha)$};
		\node at (10.75, 0) {$(1-\delta)(1-\alpha)$};
		\node at (13.25, 0) {$(1-\theta)\alpha$};
		\draw[color=black] (8.25,1.2)--(14.1,1.2);
		\draw[color=black] (9.15,1.8)--(9.15,-0.3);
		\node at (11.5,-1) {\small Alternative Plausible Prior $\pi^{\theta,\delta}$};
	\end{tikzpicture}
\end{figure} 

The parameter $\theta\in  [\underline{\theta},\overline{\theta}]$ reflects the base rate assuming the agent receives a ``correct'' signal that matches the state, while $\delta\in [\underline{\delta},\overline{\delta}]$ reflects the base rate assuming the agent receives an ``incorrect'' signal. The set of plausible priors is given by 
$$\mathbb{N}_{\theta,\delta} = \{\pi^{\theta,\delta}\mid \theta \in [\underline{\theta},\overline{\theta}], \delta\in [\underline{\delta},\overline{\delta}], \min\{\underline{\theta}, \underline{\delta}\}\geq 0, \max\{\overline{\theta}, \overline{\delta}\}\leq 1, \eta \in [\underline{\theta},\overline{\theta}]\cap [\underline{\delta},\overline{\delta}] \}.$$
As before,  $\min\{\underline{\theta}, \underline{\delta}\}\geq 0$ and $\max\{\overline{\theta}, \overline{\delta}\}\leq 1$ ensure that all the probabilities are non-negative, and  $\eta \in [\underline{\theta},\overline{\theta}]\cap [\underline{\delta},\overline{\delta}]$ ensures that $\pi\in \mathbb{N}_{\theta,\delta}$. 

Given this set of plausible priors, we can calculate the RML posteriors as
\begin{align*}
	\pi_{\sigma_1}(s_1) &= \frac{\overline{\theta}\alpha}{\overline{\theta}\alpha+(1-\underline{\delta})(1-\alpha)}, &
	\pi_{\sigma_1}(s_2) &= \frac{(1-\underline{\delta})(1-\alpha)}{\overline{\theta}\alpha+(1-\underline{\delta})(1-\alpha)}, \\[8pt]
	\pi_{\sigma_2}(s_1)  &= \frac{\overline{\delta}(1-\alpha)}{\overline{\delta}(1-\alpha)+(1-\underline{\theta})\alpha}, &
	\pi_{\sigma_2}(s_2) &= \frac{(1-\underline{\theta})\alpha}{\overline{\delta}(1-\alpha)+(1-\underline{\theta})\alpha}.
\end{align*}

From these expressions, RML updating results in base rate neglect if
$$\overline{\theta}\geq 1-\underline{\delta}, \quad \overline{\delta}\geq 1-\underline{\theta}, \quad \text{and}\quad \eta > \max \left\{\frac{\overline{\theta}}{1+\overline{\theta}-\underline{\delta}}, \frac{\overline{\delta}}{1-\underline{\theta}+\overline{\delta}}\right\}.$$
Notice that these mirror the inequalities for conservatism. If $\min\{\underline{\theta}, \underline{\delta}\}\geq 1/2$, then the first two inequalities are automatically satisfied. The requirement $\min\{\underline{\theta}, \underline{\delta}\}\geq 1/2$ is consistent with the original assumption that the agent considers $s_1$ to be more likely than $s_2$ ex ante. The last inequality implies that base rate neglect is more likely to occur if the original base rate is high. This is consistent with the findings reported in \citet{benjamin2019errors}, which show diminished sensitivity to priors that are extreme. 

Conversely, RML updating results in overconfidence if 
$$\frac{\overline{\delta}}{1-\underline{\theta}+\overline{\delta}}<\eta<\frac{\overline{\theta}}{1+\overline{\theta}-\underline{\delta}}.$$
Notice that these mirror the inequalities for confirmation bias. We can conclude that if the true base rate is high, it is more likely that the first set of inequalities will be satisfied and the agent will exhibit base rate neglect. On the other hand, if the true base rate is within the above range, the agent will display overconfidence.

\begin{figure}[htbp]
	\centering
	\includegraphics[scale=0.50]{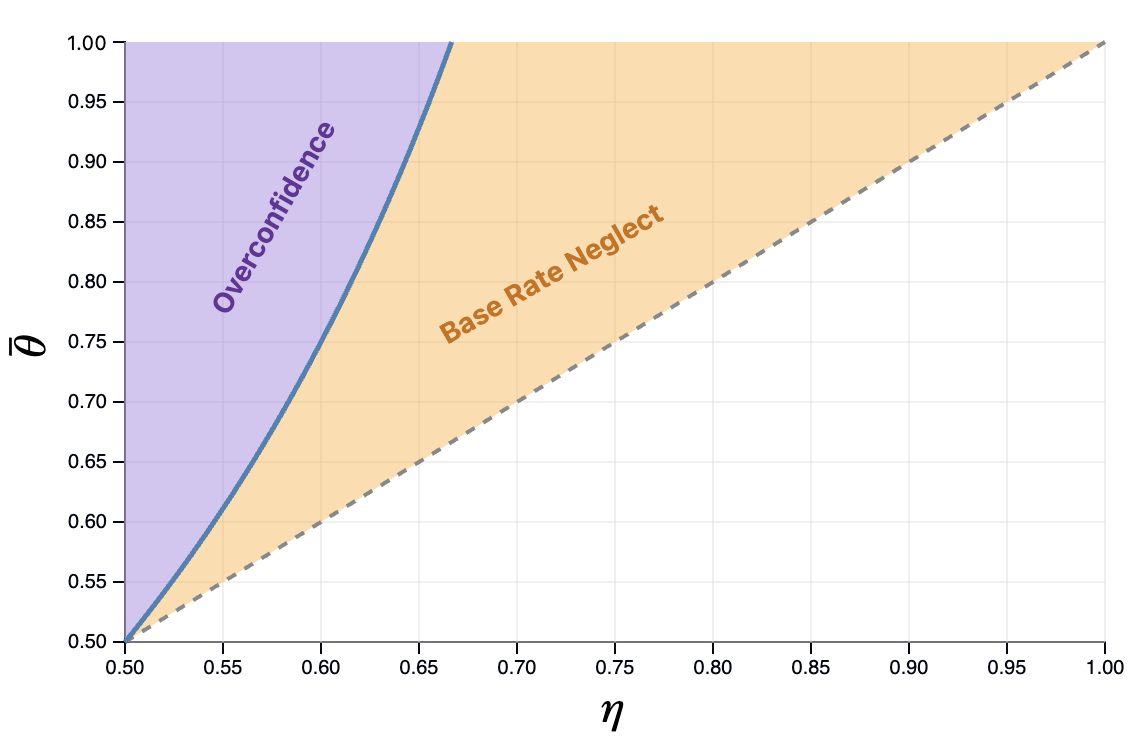}
	\caption{Parameters consistent with base rate neglect and overconfidence when $\sigma_1$ is realized, assuming $\underline{\delta}=1/2$.}
	\label{figure:BOBias}
\end{figure}

To illustrate, assume $\underline{\delta}=1/2$ and $\sigma_1$ is the realized signal. Then, base rate neglect arises if
$$\eta>\frac{2\overline{\theta}}{1+2\overline{\theta}},$$
while overconfidence implies the opposite. Figure \ref{figure:BOBias} displays the regions consistent with each bias, mirroring the relationship between confirmation bias and conservatism.

\section{Related Literature}
\label{section:lit_review}

This paper lies in the intersection of the literature on non-Bayesian updating and updating under ambiguity. The two most closely related papers are \citet{gilboa1993updating} and \citet{ortoleva2012modeling}. I first provide a discussion on the two most closely related papers and next briefly discuss other related papers.

Maximum likelihood updating was introduced by \citet{gilboa1993updating} as a dynamic extension of the maxmin expected utility model. In their model, a DM endowed with a set of priors evaluates acts according to their minimal expected utility, where the minimum is taken over all the priors in this set, and the DM performs maximum likelihood updating to revise the set of priors when she receives new information. In \citet{gilboa1993updating}, a DM whose behavior is consistent with the subjective expected utility model must follow Bayes' rule. On the other hand, an RML agent may deviate from Bayesian updating when she faces ambiguity even if she is ambiguity neutral. In addition, I show how violations of dynamic consistency can be used to uniquely identify the set of plausible priors. 

\citet{ortoleva2012modeling} axiomatizes a related updating rule, called the Hypothesis Testing (HT) model. In his model, the DM follows Bayes' rule for ``normal" events but deviates from Bayesian updating when an ``unexpected," small probability event occurs. In addition to allowing deviations from Bayesian updating, the HT model also imposes a structure on belief updating when a zero probability event occurs, which is not the case in RML updating. On the other hand, the HT model has two assumptions that are more general than RML updating: (i) the HT model imposes no structure on the set of priors the DM considers, whereas in RML updating the set of plausible priors needs to satisfy certain properties, as described in Section \ref{section:RML}, (ii) in the HT model, the agent may have any arbitrary subjective second-order prior over the set of priors, whereas in RML updating it is implicitly assumed to be uniform. Since the HT model allows for any second-order prior and it is always possible to construct a second-order prior such that the maximum likelihood stage of updating yields a unique probability, the HT model has no second stage, unlike RML updating. Overall, RML updating imposes significantly more structure on belief updating for non-zero probability events compared to the HT model. This is illustrated by the observation that when every state is non-null, as is the case in this paper, the HT model imposes almost no restriction on posteriors. In contrast, RML updating is significantly more restrictive, which allows for unique identification of plausible priors from updating behavior.

There are a few other papers that use an axiomatic approach to investigate non-Bayesian updating. \citet{epstein2006axiomatic} and \citet*{epstein2008non} axiomatize a non-Bayesian updating model where decision makers may be tempted to update their beliefs using a prior different from their original prior. In RML updating, decision makers also revise their original prior when they receive new information, but this is not due to temptation but rather due to ambiguity and willingness to make an inference. \citet{zhao2018representativeness} provides an axiomatic foundation for similarity-based updating, building on the representativeness heuristic of \citet{kahneman1972subjective}. \cite{kovach2021conservative} axiomatizes a conservative updating rule, where the DM's posterior is a mixture of her prior and Bayesian posterior, as in \citet*{epstein2010non}. While similarity-based updating, conservative updating, and RML updating can accommodate some of the same behavioral biases, the underlying behavioral motivations for these models are completely different.

Many behavioral models in the literature explain non-Bayesian updating by assuming some type of bounded rationality. This includes assuming imperfect memory \citep*{mullainathan2002memory,gennaioli2010comes, wilson2014bounded}, coarse thinking \citep*{mullainathan2002thinking,mullainathan2008coarse}, the use of the representativeness heuristic \citep*{kahneman1972subjective}, or reliance on misspecified subjective models \citep*{barberis1998model,rabin1999first} by decision makers. The main focus in these papers is providing a cognitive mechanism underlying specific updating biases, rather than characterizing an updating rule axiomatically.

The idea of distance minimization has previously been used in the literature to investigate non-standard updating behavior. \citet{zhao2022pseudo} proposes a model that allows DMs to update their beliefs when they receive new information of the form ``event $A$ is more likely than event $B$''. In his model, the posterior minimizes KL divergence from the prior subject to the informational constraint. More recently, \citet*{dominiak2025inertial} consider more general forms of information structures and distance measures. In RML updating, the idea is similar, as the DM chooses her posterior by minimizing KL divergence from the Bayesian posterior of the benchmark prior subject to the constraint that the new prior assigns the maximal likelihood to the observed event. 

A natural setup where the DM might have a benchmark prior and a set of plausible priors is when the DM receives forecasts or recommendations from different experts. \cite{levy2021maximum} study the problem of a DM who receives forecasts from multiple Bayesian forecasters and uses the maximum likelihood method to form an explanation for these forecasts. \citet*{ke2024learning} take a decision theoretic approach and study a problem where recommendations may not necessarily come from a Bayesian agent. One of the main differences in this paper is that, rather than taking the set of plausible priors as exogenous, I impose axioms on the DM's ex ante and ex post beliefs, which allows me to uniquely identify the set of plausible priors endogenously.

This paper also borrows ideas from the literature on updating under ambiguity (for a review of this literature, see \citealp{machina2014ambiguity} and \citealp{gilboa2016ambiguity}). As an alternative to maximum likelihood updating, \citet{pires2002rule} characterizes the full Bayesian updating model, where the DM updates each prior using Bayes' rule. \cite{epstein2007learning} propose a mixture model where the DM applies Bayes' rule only to a subset of priors that are considered sufficiently likely. A few other recent papers axiomatize updating rules that generalize full Bayesian and maximum likelihood updating \citep[see, for example,][]{cheng2022relative, hill2022updating, kovach2024ambiguity}. These papers do not address non-Bayesian updating when a single prior can be elicited from the DM's ex ante preferences.

Other recent papers axiomatize updating rules under ambiguity by imposing some version of dynamic consistency. \citet{epstein2003recursive} retain dynamic consistency by restricting the set of events on which the DM can update her beliefs. \citet{hanany2007updating} characterize dynamically consistent maxmin expected utility preferences without any restriction on the set of conditioning events and show that updated preferences must depend on the initial menu the DM is offered. \citet{siniscalchi2011dynamic} allows deviations from dynamic consistency but assumes that the DM can anticipate her future deviations. \citet{gul2021evaluating} consider a weaker version of dynamic consistency, which can be interpreted as ``not all news can be bad news," and show that neither maximum likelihood updating nor full Bayesian updating satisfies this property. RML updating also treats dynamic consistency as a desirable property by ensuring that the DM is ``maximally" dynamically consistent. That is, any deviation from dynamic consistency is due to the DM's willingness to use new information to make an inference on the set of plausible priors. 

Recently, a few papers investigate belief updating when agents receive ambiguous information in environments where there is not necessarily any prior ambiguity (e.g., see \citealp{shishkin2023ambiguous}, \citealp{epstein2024hard}, and \citealp{liang2025learning}). While signal ambiguity would also cause an RML agent to deviate from Bayesian updating, as demonstrated in Section \ref{section:examples}, the RML updating model involves an alternative source for non-Bayesian updating. Specifically, for an RML agent, there may also be potential prior ambiguity, which may not be reflected in ex ante preferences due to a neutral attitude towards ambiguity, but may still affect updating behavior. 

Lastly, other recent work uses ambiguity to explain deviations from Bayesian updating. \citet*{baliga2013polarization} show that belief polarization can arise when decision makers are ambiguity averse. \citet*{fryer2019updating} explain confirmation bias by assuming that when decision makers receive ambiguous signals, they interpret them as favorable to their original beliefs. In a social learning experiment, \citet*{de2022non} find that decision makers frequently deviate from Bayesian updating when they receive private information that contradicts their original beliefs. Their explanation for this phenomenon involves multiple priors. 

\section{Conclusion}
\label{section:conclusion}

This paper investigates a model where deviations from Bayesian updating arise due to the DM's subjective uncertainty regarding the accuracy of her benchmark prior. In the RML updating model, the DM uses new information to revise her benchmark prior via the maximum likelihood principle in a way that ensures maximally dynamically consistent behavior. I axiomatically characterize the updating rule, demonstrate how the DM's ex ante and ex post beliefs can be utilized to uniquely identify the set of priors she considers plausible, and show how the model can accommodate most commonly observed updating biases.

There are several potential directions for future research. First, one of the key assumptions in the paper is that the set of unambiguous events forms an algebra. It is well-known in the literature that this property may not hold for unambiguous events in general, as unambiguous events may not be closed under intersections \citep{zhang2002subjective}. A characterization of a more general model that relaxes the algebra assumption could isolate the more essential features of RML updating independent of this assumption. 

A second possible generalization would allow for more flexibility in how the DM revises the set of plausible priors when she receives new information. In the current model, the DM discards all the plausible priors that do not maximize the likelihood of the observed event. Alternatively, we could extend the model so that a plausible prior is not discarded as long as it passes a likelihood ratio test, as in \citet{epstein2007learning}. That is, we can allow that a plausible prior $\pi'$ belongs to the set $\mathbb{N}_{A}$ as long as $\pi'(A)\geq \kappa \overline{\pi}(A)$, where $\kappa \in (0,1]$. As in the current model, the DM then picks a prior from $\mathbb{N}_{A}$ that is as close to the original benchmark prior as possible and updates it using Bayes' rule. This paper only addresses the case when $\kappa = 1$. Note, however, that as $\kappa$ decreases, it becomes more and more likely that the original benchmark prior will be retained after receiving new information, and hence the updating behavior will be more in line with Bayesian updating.

A third avenue for future research is to accommodate ambiguity averse agents who evaluate acts according to their minimum expected utility over a subset of plausible priors. Specifically, the DM might be endowed with a set of plausible priors $\mathbb{N}$, representing all priors that cannot be ruled out, and a set of probable priors $\mathbb{P}$, representing all priors relevant for ex ante choices, where $\pi\in \mathbb{P}\subseteq \mathbb{N}$. This paper addresses the case where $\mathbb{P}=\{\pi\}$, while the case where $\mathbb{P}=\mathbb{N}$ coincides with the model of \cite{gilboa1993updating}. A characterization result for the intermediate case would contribute to the literature on updating under ambiguity. 

\newpage

\section*{Appendix}
\label{section:appendix}

\setcounter{equation}{0}
\renewcommand{\theequation}{A.\arabic{equation}}

\subsection*{Proof of Proposition \ref{proposition:RML}}
First, notice that if $\pi'(A) = \overline{\pi}(A)$, then it must be the case that $\pi'(A\cap P)=\overline{\pi}(A\cap P)$ for all $P\in \mathcal{P}$ with $A\cap P\neq \emptyset$. To see this, suppose not. Then, there exist $P\in \mathcal{P}$ with $A\cap P\neq \emptyset$ and $\pi''\in \mathbb{N}$ such that $\pi''(A\cap P)>\pi'(A\cap P)$. Construct another prior $\pi'''$ by setting
$$\pi'''|_{P} = \pi''|_{P} \quad \text{and} \quad \pi'''|_{P^c}=\pi'|_{P^c}.$$
By the $\mathcal{P}$-rectangularity property of $\mathbb{N}$, we have $\pi'''\in \mathbb{N}$. But then $\pi'''(A)>\pi'(A)$, contradicting $\pi'(A) = \overline{\pi}(A)$. Hence, we must have $\pi'(A\cap P)=\overline{\pi}(A\cap P)$ for all $P\in \mathcal{P}$ with $A\cap P\neq \emptyset$.

Next, notice that if $\pi'_A\in B(\mathbb{N}_A)$, then by the previous argument, we must have 
\begin{equation}
	\label{equation:KL.3}
	\sum_{\omega\in A\cap P}\pi'_A(\omega)=\frac{\overline{\pi}(A\cap P)}{\sum_{P'\in \mathcal{P}:\: A\cap P'\neq \emptyset}\overline{\pi}(A\cap P')}
\end{equation}
for all $P\in \mathcal{P}$ with $A\cap P\neq \emptyset$. Hence, equation \ref{equation:KL.3} is a constraint in the minimization problem in Definition \ref{definition:RML}. However, this is not necessarily the only constraint, as it is possible that $\pi'_A$ satisfies equation \ref{equation:KL.3} but nevertheless $\pi'_A\notin B(\mathbb{N}_A)$. 

Consider a relaxed problem where equation \ref{equation:KL.3} is the only constraint. The objective in the relaxed problem is to minimize $D_{\text{KL}}(\pi(\cdot|A)\: ||\: \pi_A')$ subject to the constraints given by equation \ref{equation:KL.3} for each $P\in \mathcal{P}$ with $A\cap P\neq \emptyset$. The Lagrangian $\mathcal{L}\left(\{\pi_A'(\omega)\}_{\omega\in A}, \{\lambda_{P}\}_{P\in \mathcal{P}: A\cap P\neq \emptyset}\right)$ for the relaxed problem is given by
{\small
	$$-\sum_{\omega\in A}\pi(\omega|A)\ln\left(\frac{\pi_A'(\omega)}{\pi(\omega|A)}\right)
	+\sum_{P\in \mathcal{P}:A\cap P\neq \emptyset}\lambda_{P}\left(\sum_{\omega\in A\cap P}\pi_A'(\omega)-\frac{\overline{\pi}(A\cap P)}{\sum_{P'\in \mathcal{P}:\: A\cap P'\neq \emptyset}\overline{\pi}(A\cap P')}\right).$$
}The first-order conditions imply that for any $P\in \mathcal{P}$ with $A\cap P\neq \emptyset$ and any $\omega,\omega'\in A\cap P$,
\begin{align}
	\label{equation:KL.4}
	\frac{\pi(\omega|A)}{\pi'_A(\omega)}=\lambda_{P}=\frac{\pi(\omega'|A)}{\pi'_A(\omega')}, \quad \text{and hence} \quad  \frac{\pi_A'(\omega)}{\pi_A'(\omega')}=\frac{\pi(\omega|A)}{\pi(\omega'|A)}=\frac{\pi(\omega)}{\pi(\omega')}.
\end{align}
Hence, equations \ref{equation:KL.3} and \ref{equation:KL.4} characterize the solution to the relaxed minimization problem. Now, by the prior-coherence property of $\mathbb{N}$, there exists a prior $\pi''\in \mathbb{N}$ that maximizes the likelihood of $A$ and preserves the relative likelihood of any two states within minimal unambiguous events. Thus, $\pi''(\cdot|A)$ satisfies both equations \ref{equation:KL.3} and \ref{equation:KL.4}, and hence it solves the relaxed problem. Since $\pi''(\cdot|A)$ solves the relaxed problem (with equation \ref{equation:KL.3} as the only constraint) and belongs to the original constraint set, it must also solve the original optimization problem in Definition \ref{definition:RML}. Therefore, equations \ref{equation:KL.3} and \ref{equation:KL.4} fully characterize the solution to the problem in Definition \ref{definition:RML}. Combining these two equations, we conclude that the RML posterior is given by equation \ref{equation:two_stage2}. $\hfill\square$

\subsection*{Proof of Proposition \ref{proposition:ideal}}

By the previous proposition, we can use the representation for RML updating given by equation \ref{equation:two_stage2}. If $\mathcal{P} = \{\Omega\}$, then every event is ideal, and updating coincides with Bayesian updating. Therefore, this case cannot be distinguished from the case when every event is unambiguous. Hence, we can assume $\mathcal{P}\neq \{\Omega\}$.

First, suppose $E$ is an unambiguous event (i.e., $E\in \sigma(\mathcal{P})$). Let $A\supseteq E$ be given. Then, for any $\omega\in E$,
\begin{align*}
	\pi_A(\omega) & = \frac{\pi(\omega)}{\sum_{\omega'\in A\cap P_{\omega}}\pi(\omega')} \cdot \frac{\overline{\pi}(A\cap P_\omega)}{\sum_{P\in \mathcal{P}: A\cap P\neq \emptyset}\overline{\pi}(A\cap P)}\\
	& = \frac{\pi(\omega)}{\sum_{\omega'\in P_{\omega}}\pi(\omega')} \cdot \frac{\overline{\pi}(P_\omega)}{\sum_{P\in \mathcal{P}: A\cap P\neq \emptyset}\overline{\pi}(A\cap P)}\\
	& = \frac{\pi(\omega)}{\pi(P_{\omega})} \cdot \frac{\pi(P_\omega)}{\sum_{P\in \mathcal{P}: A\cap P\neq \emptyset}\overline{\pi}(A\cap P)}\\
	& = \frac{\pi(\omega)}{\sum_{P\in \mathcal{P}: A\cap P\neq \emptyset}\overline{\pi}(A\cap P)},
\end{align*}
where the second equality is due to $P_{\omega}\subseteq E\subseteq A$ (since $E\in \sigma(\mathcal{P})$), and the third equality follows from $\overline{\pi}(P_\omega) = \pi(P_\omega)$. Thus, for any $\omega,\omega'\in E$ and $A\supseteq E$,
\begin{align*}
	\frac{\pi_A(\omega)}{\pi_A(\omega')} = \frac{\pi(\omega)}{\pi(\omega')} =  \frac{\pi_E(\omega)}{\pi_E(\omega')}. 
\end{align*}
Therefore, $E$ is an ideal event. Similarly, $E^c$ is also ideal.

Now, suppose $E$ is an ambiguous event (i.e., $E\notin \sigma(\mathcal{P})$). It needs to be shown that either $E$ or $E^c$ is not ideal. There are two cases to consider.

First, suppose there exists $P\in \mathcal{P}$ such that $E\subsetneq P$. Let $A=E^c= (P\setminus E)\cup P^c$. By the non-extremeness property of $\mathbb{N}$, $\overline{\pi}(P\setminus E)> \pi(P\setminus E)$. Additionally, for any $\omega\in P\setminus E$, we have $A\cap P_{\omega}=P\setminus E$. Thus, 
\begin{align*}
	\pi_A(\omega) & = \frac{\pi(\omega)}{\sum_{\omega''\in A\cap P_{\omega}}\pi(\omega'')} \cdot \frac{\overline{\pi}(A\cap P_\omega)}{\sum_{P'\in \mathcal{P}: A\cap P'\neq \emptyset}\overline{\pi}(A\cap P')}\\
	&= \frac{\pi(\omega)}{\pi(P\setminus E)} \cdot \frac{\overline{\pi}(P\setminus E)}{\sum_{P'\in \mathcal{P}: A\cap P'\neq \emptyset}\overline{\pi}(A\cap P')}.
\end{align*}
On the other hand, for $\omega'\in P^c$,
\begin{align*}
	\pi_A(\omega') & = \frac{\pi(\omega')}{\sum_{\omega''\in A\cap P_{\omega'}}\pi(\omega'')} \cdot \frac{\overline{\pi}(A\cap P_{\omega'})}{\sum_{P'\in \mathcal{P}: A\cap P'\neq \emptyset}\overline{\pi}(A\cap P')} \\
	& = \frac{\pi(\omega')}{\pi(P_{\omega'})} \cdot \frac{\pi(P_{\omega'})}{\sum_{P'\in \mathcal{P}: A\cap P'\neq \emptyset}\overline{\pi}(A\cap P')} \\
	&= \frac{\pi(\omega')}{\sum_{P'\in \mathcal{P}: A\cap P'\neq \emptyset}\overline{\pi}(A\cap P')},
\end{align*}
where the second equality follows from $P^c \subseteq A$ and the fact that $P_{\omega'}$ is unambiguous (implying $\overline{\pi}(A\cap P_{\omega'})=\overline{\pi}(P_{\omega'})=\pi(P_{\omega'})$). Therefore, 
$$\frac{\pi_A(\omega)}{\pi_A(\omega')} = \frac{\pi(\omega)}{\pi(\omega')} \cdot \frac{\overline{\pi}(P\setminus E)}{\pi(P\setminus E)}>\frac{\pi(\omega)}{\pi(\omega')}.$$
Hence, $E^c = A$ is not ideal.

Next, suppose there are $P,P'\in \mathcal{P}$ such that $E\cap P\neq \emptyset$, $E\cap P'\neq \emptyset$, and either $P\setminus E\neq \emptyset$ or $P'\setminus E\neq \emptyset$. Without loss of generality, assume $P\setminus E\neq \emptyset$. Consider $A=E\cup P'$. Let $\omega\in E\cap P$ and $\omega' \in E\cap P'$ be given. We have
\begin{align*}
	\pi_A(\omega) & = \frac{\pi(\omega)}{\sum_{\omega''\in A\cap P_{\omega}}\pi(\omega'')} \cdot \frac{\overline{\pi}(A\cap P_{\omega})}{\sum_{P''\in \mathcal{P}: A\cap P''\neq \emptyset}\overline{\pi}(A\cap P'')}\\
	& = \frac{\pi(\omega)}{\pi(A\cap P)} \cdot \frac{\overline{\pi}(A\cap P)}{\sum_{P''\in \mathcal{P}: A\cap P''\neq \emptyset}\overline{\pi}(A\cap P'')},
\end{align*}
where $\pi(A\cap P) < \overline{\pi}(A\cap P)$ since $P\setminus E = P\setminus A \neq \emptyset$. On the other hand,
\begin{align*}
	\pi_A(\omega') & = \frac{\pi(\omega')}{\sum_{\omega''\in A\cap P_{\omega'}}\pi(\omega'')} \cdot \frac{\overline{\pi}(A\cap P_{\omega'})}{\sum_{P''\in \mathcal{P}: A\cap P''\neq \emptyset}\overline{\pi}(A\cap P'')} \\
	& = \frac{\pi(\omega')}{\pi(P')} \cdot \frac{\overline{\pi}(P')}{\sum_{P''\in \mathcal{P}: A\cap P''\neq \emptyset}\overline{\pi}(A\cap P'')}\\
	& = \frac{\pi(\omega')}{\sum_{P''\in \mathcal{P}: A\cap P''\neq \emptyset}\overline{\pi}(A\cap P'')},
\end{align*}
where the second equality is due to $A\cap P' = P'$, and the last equality holds because $P'$ is unambiguous. Thus, 
$$\frac{\pi_A(\omega)}{\pi_A(\omega')} = \frac{\pi(\omega)}{\pi(\omega')} \cdot \frac{\overline{\pi}(A\cap P)}{\pi(A\cap P)}>\frac{\pi(\omega)}{\pi(\omega')}.$$
This shows $E$ is not ideal, as $\pi_E$ cannot simultaneously be the Bayesian update of both $\pi_A$ and $\pi$ conditional on the event $E$. $\hfill\square$

\subsection*{Proof of Theorem \ref{theorem:RML}}

\subsubsection*{Necessity}
The necessity of Axiom \ref{axiom:regularity} follows from the assumption that $\pi$ has full support and the representation for RML updating given by equation \ref{equation:two_stage2}. For the remaining axioms, first note that if $\mathcal{P}=\{\Omega\}$, then updating is Bayesian and all the axioms trivially hold. Hence, assume $\mathcal{P}\neq \{\Omega\}$. Axiom \ref{axiom:algebra2} then directly follows from Proposition \ref{proposition:ideal}, which shows that the collection of events that are symmetrically ideal corresponds to the set of unambiguous events under RML updating. Hence, $\mathcal{I} = \sigma(\mathcal{P})$ and $\mathcal{P} = \mathcal{P}_{\mathcal{I}}$. Axiom \ref{axiom:ideal} follows from the representation for RML updating given by equation \ref{equation:two_stage2}, as it implies that for any $D\subsetneq P$, where $P\in \mathcal{P}$, $A\supseteq D$, and $\omega,\omega'\in D$, we have 
$$ \frac{\pi_D(\omega)}{\pi_D(\omega')}= \frac{\pi_A(\omega)}{\pi_A(\omega')}.$$ 
For Axiom \ref{axiom:btw}, notice that 
\begin{align*}
	\pi_A(P) & = \frac{\overline{\pi}(A\cap P)}{\overline{\pi}(A\cap P) + \sum_{P'\in \mathcal{P}\setminus P: A\cap P'\neq \emptyset}\overline{\pi}(A\cap P')} \\
	& \geq \frac{\overline{\pi}((A\setminus D)\cap P)}{\overline{\pi}((A\setminus D)\cap P) + \sum_{P'\in \mathcal{P}\setminus P: A\cap P'\neq \emptyset}\overline{\pi}(A\cap P')}\\
	& = \frac{\overline{\pi}((A\setminus D)\cap P)}{\overline{\pi}((A\setminus D)\cap P) + \sum_{P'\in \mathcal{P}\setminus P: (A\setminus D)\cap P'\neq \emptyset}\overline{\pi}((A\setminus D)\cap P')}\\
	& = \pi_{A\setminus D}(P),
\end{align*}
where the first and the last equalities are due to the representation in equation \ref{equation:two_stage2}, the inequality is due to $\overline{\pi}(A\cap P)\geq \overline{\pi}((A\setminus D)\cap P)$, and the second equality is due to $D\subsetneq P$ which implies $A\cap P'=(A\setminus D)\cap P'$ for all $P'\neq P$.

For the second part of the axiom, notice that $\pi_A$ is the Bayesian posterior of some prior $\pi'\in \mathbb{N}$ that satisfies $\pi'(A\cap P') = \overline{\pi}(A\cap P')$ for all $P'\in \mathcal{P}$ with $A\cap P'\neq \emptyset$. Hence, $\pi_A(\cdot|A\setminus D)$ is also the Bayesian posterior of $\pi'$ conditional on the event $A\setminus D$. Therefore,
\begin{align*}
	\pi_A(P|A\setminus D) & = \frac{\pi'((A\setminus D)\cap P)}{\pi'((A\setminus D)\cap P) + \sum_{P'\in \mathcal{P}\setminus P: (A\setminus D) \cap P'\neq \emptyset}\pi'((A\setminus D)\cap P')}\\
	& = \frac{\pi'((A\setminus D)\cap P)}{\pi'((A\setminus D)\cap P) + \sum_{P'\in \mathcal{P}\setminus P: A\cap P'\neq \emptyset}\pi'(A\cap P')}\\
	& =  \frac{\pi'((A\setminus D)\cap P)}{\pi'((A\setminus D)\cap P) + \sum_{P'\in \mathcal{P}\setminus P: A\cap P'\neq \emptyset}\overline{\pi}(A\cap P')}\\
	& \leq \frac{\overline{\pi}((A\setminus D)\cap P)}{\overline{\pi}((A\setminus D)\cap P) + \sum_{P'\in \mathcal{P}\setminus P: A\cap P'\neq \emptyset}\overline{\pi}(A\cap P')}\\
	& = \pi_{A\setminus D}(P),
\end{align*}
where the first equality is due to the argument before the expression, the second equality is due to $D\subsetneq P$, the third equality is due to $\pi'(A\cap P') = \overline{\pi}(A\cap P')$ for all $P'$ with $A\cap P'\neq \emptyset$, the inequality is due to $\pi'\in \mathbb{N}$ which implies $\pi'((A\setminus D)\cap P)\leq \overline{\pi}((A\setminus D)\cap P)$, and the last equality is due to the representation for RML updating given by equation \ref{equation:two_stage2}.

For Axiom \ref{axiom:sbm}, first note that if the updating rule is consistent with RML updating, then 
$$O_P(A) = \frac{\pi_A(P)}{1-\pi_A(P)}= \frac{\overline{\pi}(A\cap P)}{\sum_{P'\in \mathcal{P}\setminus P: A\cap P'\neq \emptyset}\overline{\pi}(A\cap P')}.$$
Let $D_1,D_2\subseteq P$ be such that $D_1\cap D_2=\emptyset$. Notice that for any $P'\in \mathcal{P}\setminus P$, $A\cap P' = (A\setminus D_1) \cap P' = (A\setminus D_2) \cap P'= (A\setminus (D_1\cup D_2)) \cap P'$. Therefore, the denominators in all the expressions $O_P(A)$, $O_P(A\setminus D_1)$, $O_P(A\setminus D_2)$, $O_P(A\setminus (D_1\cup D_2))$ are the same. Hence, the submodularity property is equivalent to 
$$\overline{\pi}(A\cap P) + \overline{\pi}((A\setminus (D_1\cup D_2))\cap P) \leq \overline{\pi}((A\setminus D_1 )\cap P) + \overline{\pi}((A\setminus D_2)\cap P).$$
This is implied by the chain-coherence property of $\mathbb{N}$. To see this, consider the chain $(A\setminus (D_1\cup D_2))\cap P\subseteq (A\setminus D_1 )\cap P \subseteq A\cap P$ within $P$. By chain-coherence, there exists $\pi'\in \mathbb{N}$ such that $\pi'$ maximizes the likelihood of all three events within $\mathbb{N}$. In addition, by definition, we must have $\pi'((A\setminus D_2)\cap P)\leq \overline{\pi}((A\setminus D_2)\cap P)$. Therefore, 
\begin{align*}
	\overline{\pi}(A\cap P) + \overline{\pi}((A\setminus (D_1\cup D_2))\cap P)  = \pi'(A\cap P) + \pi'((A\setminus (D_1\cup D_2))\cap P)  \\
	= \pi'((A\setminus D_1 )\cap P) + \pi'((A\setminus D_2)\cap P) \\
	\leq \overline{\pi}((A\setminus D_1 )\cap P) + \overline{\pi}((A\setminus D_2)\cap P),
\end{align*}
as desired.

For Axiom \ref{axiom:lmd}, let $P\in \mathcal{P}$, $A\in \mathcal{A}$ with $A\cap P\neq \emptyset$, $A\cap P^c\neq \emptyset$, $D_1\subsetneq A\cap P$, and $D_2\subsetneq A\cap P^c$ be given. First, notice that, by the $\mathcal{P}$-rectangularity property of $\mathbb{N}$, $\sum_{P'\in \mathcal{P}\setminus P: A\cap P'\neq \emptyset}\overline{\pi}(A\cap P') = \overline{\pi}(A\cap P^c)$. Hence, 
\begin{align*}
	O_P(A)O_P(A\setminus (D_1\cup D_2)) &=  \frac{\overline{\pi}(A\cap P)}{\overline{\pi}(A\cap P^c)}\frac{\overline{\pi}((A\setminus (D_1\cup D_2))\cap P)}{\overline{\pi}((A\setminus (D_1\cup D_2))\cap P^c)} \\
	& = \frac{\overline{\pi}(A\cap P)}{\overline{\pi}(A\cap P^c)}\frac{\overline{\pi}((A\setminus D_1)\cap P)}{\overline{\pi}((A\setminus D_2)\cap P^c)} \\
	&  = \frac{\overline{\pi}((A\setminus D_1)\cap P)}{\overline{\pi}((A\setminus D_1)\cap P^c)}\frac{\overline{\pi}((A\setminus D_2)\cap P)}{\overline{\pi}((A\setminus D_2)\cap P^c)}\\
	& = O_P(A\setminus D_1)O_P(A\setminus D_2),
\end{align*}
where the second and the third equalities follow from $D_1\subsetneq A\cap P$ and $D_2\subsetneq A\cap P^c$. The above equality is equivalent to log-modularity.

\subsubsection*{Sufficiency}

By Axiom \ref{axiom:algebra2}, the collection of symmetrically ideal events forms an algebra and is generated by the partition of the state space $\mathcal{P}_{\mathcal{I}}$. If $\mathcal{P}_{\mathcal{I}} = \{\Omega\}$, then every event is ideal and hence $\{\pi_A\}_{A\in \mathcal{A}}$ is consistent with RML updating with $\mathbb{N}=\{\pi\}$. Hence, we can assume $\mathcal{P}_{\mathcal{I}}$ is not degenerate and let $\mathcal{P}_{\mathcal{I}} = \{P_1,\ldots, P_n\}$. 

We next construct the set of plausible priors $\mathbb{N}$. First, for each $P_i\in \mathcal{P}_{\mathcal{I}}$, we construct the set $\mathbb{N}(P_i)$ as follows. Enumerate the states in $P_i$ as $\{\omega_{i1},\dots, \omega_{ik_i}\}$. For any $D\subseteq P_i$, let $\mu_{D,P_i}:\{1,2,\ldots,k_i\}\rightarrow \{1,2,\ldots,k_i\}$ denote a permutation such that if $\omega_{il}\in D$ and $\omega_{im}\in P_i\setminus D$, then $\mu_{D,P_i}(l) < \mu_{D,P_i}(m)$. Let $\mathcal{M}_{D,P_i}$ denote all such permutations. Next, for each $D\subseteq P_i$ and $\mu_{D,P_i}\in \mathcal{M}_{D,P_i}$, we will construct a prior $\pi^{\mu_{D,P_i}}$ that assigns the highest probability to $D$ among all plausible priors. First, if $\omega_{ij}\in D$, then 
$$\pi^{\mu_{D,P_i}}(\omega_{ij})=\frac{\pi_{P_i^c\cup D}(P_i)\pi(P_i^c)\pi(\omega_{ij})}{\pi_{P_i^c\cup D}(P_i^c)\pi(D)}.$$ 
Axiom \ref{axiom:regularity} guarantees that the denominator in the above expression is strictly positive and that $\pi^{\mu_{D,P_i}}(\omega_{ij})>0$. In addition, $\pi_{P_i^c\cup D}(P_i)=\pi_{P_i^c\cup D}(D)$ must hold, so we can use either term interchangeably. Notice that the priors $\pi^{\mu_{D,P_i}}$ agree on $D$ for all $\mu_{D,P_i}\in \mathcal{M}_{D,P_i}$. 

Next, given $D\subseteq P_i$ and $\mu_{D,P_i}\in \mathcal{M}_{D,P_i}$, let $D_{\mu_j}=\{\omega_{im}\in P_i \mid \mu_{D,P_i}(m)\leq \mu_{D,P_i}(j)\}$. Notice that $D\subseteq D_{\mu_j}$ if $\omega_{ij}\in P_i\setminus D$. Now, if $\omega_{ij}\in P_i\setminus D$, define $\pi^{\mu_{D,P_i}}(\omega_{ij})$ as 
$$\pi^{\mu_{D,P_i}}(\omega_{ij})=\frac{\pi_{P_i^c\cup D_{\mu_j}}(P_i)\pi(P_i^c)}{\pi_{P_i^c\cup D_{\mu_j}}(P_i^c)}-\frac{\pi_{P_i^c\cup D_{\mu_j}\setminus \{\omega_{ij}\}}(P_i)\pi(P_i^c)}{\pi_{P_i^c\cup D_{\mu_j}\setminus \{\omega_{ij}\}}(P_i^c)}.$$
Since Axiom \ref{axiom:btw} implies 
$$\frac{\pi_{P_i^c\cup D_{\mu_j}}(P_i)}{\pi_{P_i^c\cup D_{\mu_j}}(P_i^c)}-\frac{\pi_{P_i^c\cup D_{\mu_j}\setminus \{\omega_{ij}\}}(P_i)}{\pi_{P_i^c\cup D_{\mu_j}\setminus \{\omega_{ij}\}}(P_i^c)}\geq 0,$$
$\pi^{\mu_{D,P_i}}(\omega_{ij})\geq 0$ must hold for all $\omega_{ij}\in P_i$. In addition, notice that
{\small
	\begin{align*}
		\pi^{\mu_{D,P_i}}(P_i) & = \pi^{\mu_{D,P_i}}(D) +  \sum_{\omega_{ij}\in P_i\setminus D}\pi^{\mu_{D,P_i}}(\omega_{ij}) \\
		& = \frac{\pi_{P_i^c\cup D}(P_i)\pi(P_i^c)}{\pi_{P_i^c\cup D}(P_i^c)} + \sum_{\omega_{ij}\in P_i\setminus D} \left(\frac{\pi_{P_i^c\cup D_{\mu_j}}(P_i)\pi(P_i^c)}{\pi_{P_i^c\cup D_{\mu_j}}(P_i^c)}-\frac{\pi_{P_i^c\cup D_{\mu_j}\setminus \{\omega_{ij}\}}(P_i)\pi(P_i^c)}{\pi_{P_i^c\cup D_{\mu_j}\setminus \{\omega_{ij}\}}(P_i^c)}\right)\\
		& =\frac{\pi(P_i)\pi(P_i^c)}{\pi(P_i^c)} = \pi(P_i).
	\end{align*}
}Since $\pi^{\mu_{D,P_i}}$ and $\pi$ agree on the likelihood of $P_i$, we can let $\pi^{\mu_{D,P_i}}(\omega)=\pi(\omega)$ for all states $\omega\notin P_i$. Therefore, by construction, $\pi^{\mu_{D,P_i}}$ and $\pi$ agree on the likelihoods of all unambiguous events. 

Lastly, let $\mathbb{N}(P_i)=\{\pi^{\mu_{D,P_i}}\mid D\subseteq P_i \text{ and } \mu_{D,P_i}\in \mathcal{M}_{D,P_i}\}$ and 
\begin{equation}
	\label{equation:priors}
	\mathbb{N} = \{\pi'\in \Delta(\Omega) \mid \pi'|_{P_i} = \text{co}\left(\{\pi''|_{P_i}\}_{\pi''\in \mathbb{N}(P_i)}\right)  \text{ for all } P_i\in \mathcal{P}_{\mathcal{I}}\},
\end{equation}
where $\pi'|_{P_i}$ denotes the restriction of $\pi'$ to $P_i$ and $\text{co}(\cdot)$ represents the convex hull operator. That is, $\pi'$ is a plausible prior if  for each $P_i\in \mathcal{P}_{\mathcal{I}}$, $\pi'|_{P_i}$ is in the convex hull of $\pi^{\mu_{D,P_i}}|_{P_i}$ for $D\subseteq P_i$ and $\mu_{D,P_i}\in \mathcal{M}_{D,P_i}$. 

For each $D\subseteq P_i$, let $\pi^D$ denote the prior in $\mathbb{N}$ such that $\pi^D=\pi^{\mu_{D,P_i}}$ where $\mu_{D,P_i}$ preserves the ordering of elements in $D$ and $P_i\setminus D$, i.e., if $\omega_{il},\omega_{im}\in D$ with $l<m$, then $\mu_{D,P_i}(l)<\mu_{D,P_i}(m)$ and similarly for $P_i\setminus D$. The next claim shows that $\pi^D$ attains the highest probability for $D$ among all the priors in $\mathbb{N}$. Since $\pi^D|_{D}=\pi^{\mu_{D,P_i}}|_{D}$ for all $\mu_{D,P_i}\in \mathcal{M}_{D,P_i}$, this also shows the same result for all $\pi^{\mu_{D,P_i}}$.\footnote{It will become clear later that the reason for considering all possible permutations $\mu_{D,P_i}$ is to ensure that $\mathbb{N}$ satisfies the chain-coherence property. As shown in Claim \ref{claim:uniqueness}, $\mathbb{N}$ is the unique rich set of plausible priors consistent with RML updating. The claim also provides a simpler algorithm to construct $\mathbb{N}$, as illustrated in Example \ref{example:RML}.}

\begin{claim}
	\label{claim:ML}
	For any $P_i \in \mathcal{P}_{\mathcal{I}}$ and  $D \subseteq P_i$, the prior $\pi^{D}$ attains the maximum probability for $D$ among all the priors in $\mathbb{N}$. 
\end{claim}

\begin{proof}
	
	By construction, it suffices to show that $\pi^D$ achieves the highest probability for $D$ in $\mathbb{N}(P_i)$, i.e., $\pi^D(D)\geq \pi^{\mu_{D',P_i}}(D)$ for any $D'\subseteq P_i$ and $\mu_{D',P_i}\in \mathcal{M}_{D',P_i}$. The case when $D=P_i$ is trivial, as $\pi^{P_i}(P_i)=\pi(P_i) =\pi^{\mu_{D',P_i}}(P_i)$. Hence, let $D\subsetneq P_i$. There are a few cases to consider.
	
	\noindent\textbf{Case 1:} $D'\supseteq D.$ By definition of $\pi^{\mu_{D',P_i}}$,
	$$\pi^{\mu_{D',P_i}}(D)=\frac{\pi_{P_i^c\cup D'}(P_i)\pi(P_i^c)\pi(D)}{\pi_{P_i^c\cup D'}(P_i^c)\pi(D')}.$$
	In addition, by Axiom \ref{axiom:ideal},
	$$\pi_{P_i^c\cup D'}(D)=\pi_{P_i^c\cup D'}(P_i)\frac{\pi(D)}{\pi(D')}.$$
	Therefore, 
	$$\frac{\pi_{P_i^c\cup D'}(P_i|P_i^c\cup D)}{\pi_{P_i^c\cup D'}(P_i^c|P_i^c\cup D)}
	=\frac{\pi_{P_i^c\cup D'}(D)}{\pi_{P_i^c\cup D'}(P_i^c)}
	=\frac{\pi_{P_i^c\cup D'}(P_i)\pi(D)/\pi(D')}{\pi_{P_i^c\cup D'}(P_i^c)}
	=\frac{\pi_{P_i^c\cup D'}(P_i)\pi(D)}{\pi_{P_i^c\cup D'}(P_i^c)\pi(D')}.$$
	By Axiom \ref{axiom:btw},
	$$\frac{\pi_{P_i^c\cup D'}(P_i|P_i^c\cup D)}{\pi_{P_i^c\cup D'}(P_i^c|P_i^c\cup D)}\leq \frac{\pi_{P_i^c\cup D}(P_i)}{\pi_{P_i^c\cup D}(P_i^c)}.$$
	Combining all the expressions, we get
	\begin{align*}
		\pi^{\mu_{D',P_i}}(D)
		&=\frac{\pi_{P_i^c\cup D'}(P_i)\pi(P_i^c)\pi(D)}{\pi_{P_i^c\cup D'}(P_i^c)\pi(D')}
		=\frac{\pi_{P_i^c\cup D'}(P_i|P_i^c\cup D)\pi(P_i^c)}{\pi_{P_i^c\cup D'}(P_i^c|P_i^c\cup D)}\\
		&\leq \frac{\pi_{P_i^c\cup D}(P_i)\pi(P_i^c)}{\pi_{P_i^c\cup D}(P_i^c)}=\pi^D(D).
	\end{align*}
	
	\noindent\textbf{Case 2:} $D'\subseteq D$. Let $D'_{\mu_j}=\{\omega_{im}\in P_i\mid \mu_{D',P_i}(m)\leq \mu_{D',P_i}(j)\}$ as before and note that $D'\subseteq D'_{\mu_j}$ if $\omega_{ij}\notin D'$. By definition, 
	{\small
		$$\pi^{\mu_{D',P_i}}(D)=\frac{\pi_{P_i^c\cup D'}(P_i)\pi(P_i^c)}{\pi_{P_i^c\cup D'}(P_i^c)}+\sum_{\omega_{ij}\in D\setminus D'}\left(\frac{\pi_{P_i^c\cup D'_{\mu_j}}(P_i)\pi(P_i^c)}{\pi_{P_i^c\cup D'_{\mu_j}}(P_i^c)}-\frac{\pi_{P_i^c\cup D'_{\mu_j}\setminus \{\omega_{ij}\}}(P_i)\pi(P_i^c)}{\pi_{P_i^c\cup D'_{\mu_j}\setminus \{\omega_{ij}\}}(P_i^c)}\right).$$
	}
	Since $(P_i\setminus D) \cap \{\omega_{ij}\}=\emptyset$ for each $\omega_{ij}\in D\setminus D'$, by Axiom \ref{axiom:sbm},
	\begin{multline*}
		\frac{\pi_{P_i^c\cup D'_{\mu_j}}(P_i)\pi(P_i^c)}{\pi_{P_i^c\cup D'_{\mu_j}}(P_i^c)}
		+\frac{\pi_{(P_i^c\cup D'_{\mu_j})\setminus ((P_i\setminus D) \cup \{\omega_{ij}\})}(P_i)\pi(P_i^c)}{\pi_{(P_i^c\cup D'_{\mu_j})\setminus ((P_i\setminus D) \cup \{\omega_{ij}\})}(P_i^c)}  \leq \\
		\frac{\pi_{(P_i^c\cup D'_{\mu_j})\setminus (P_i\setminus D)}(P_i)\pi(P_i^c)}{\pi_{(P_i^c\cup D'_{\mu_j})\setminus (P_i\setminus D)}(P_i^c)}
		+ \frac{\pi_{P_i^c\cup D'_{\mu_j}\setminus \{\omega_{ij}\}}(P_i)\pi(P_i^c)}{\pi_{P_i^c\cup D'_{\mu_j}\setminus \{\omega_{ij}\}}(P_i^c)},
	\end{multline*}
	which implies 
	\begin{multline*}
		\frac{\pi_{P_i^c\cup D'_{\mu_j}}(P_i)\pi(P_i^c)}{\pi_{P_i^c\cup D'_{\mu_j}}(P_i^c)}
		- \frac{\pi_{P_i^c\cup D'_{\mu_j}\setminus \{\omega_{ij}\}}(P_i)\pi(P_i^c)}{\pi_{P_i^c\cup D'_{\mu_j}\setminus \{\omega_{ij}\}}(P_i^c)} \leq \\
		\frac{\pi_{(P_i^c\cup D'_{\mu_j})\setminus (P_i\setminus D)}(P_i)\pi(P_i^c)}{\pi_{(P_i^c\cup D'_{\mu_j})\setminus (P_i\setminus D)}(P_i^c)}
		- \frac{\pi_{(P_i^c\cup D'_{\mu_j})\setminus ((P_i\setminus D) \cup \{\omega_{ij}\})}(P_i)\pi(P_i^c)}{\pi_{(P_i^c\cup D'_{\mu_j})\setminus ((P_i\setminus D) \cup \{\omega_{ij}\})}(P_i^c)}.
	\end{multline*}
	Therefore, 
	{\small
		\begin{align*}
			\pi^{\mu_{D',P_i}}(D) \leq & \frac{\pi_{P_i^c\cup D'}(P_i)\pi(P_i^c)}{\pi_{P_i^c\cup D'}(P_i^c)}\\
			+ &\sum_{\omega_{ij}\in D\setminus D'}\left(\frac{\pi_{(P_i^c\cup D'_{\mu_j})\setminus (P_i\setminus D)}(P_i)\pi(P_i^c)}{\pi_{(P_i^c\cup D'_{\mu_j})\setminus (P_i\setminus D)}(P_i^c)}-\frac{\pi_{(P_i^c\cup D'_{\mu_j})\setminus ((P_i\setminus D) \cup \{\omega_{ij}\})}(P_i)\pi(P_i^c)}{\pi_{(P_i^c\cup D'_{\mu_j})\setminus ((P_i\setminus D) \cup \{\omega_{ij}\})}(P_i^c)}\right) \\
			= & \frac{\pi_{P_i^c\cup D}(P_i)\pi(P_i^c)}{\pi_{P_i^c\cup D}(P_i^c)} = \pi^D(D).
		\end{align*}
	}
	
	\noindent\textbf{Case 3:} $D'\cap D = \emptyset$. By definition, 
	$$\pi^{\mu_{D',P_i}}(D)
	=\sum_{\omega_{ij}\in D}\left(\frac{\pi_{P_i^c\cup D'_{\mu_j}}(P_i)\pi(P_i^c)}{\pi_{P_i^c\cup D'_{\mu_j}}(P_i^c)}-\frac{\pi_{P_i^c\cup D'_{\mu_j}\setminus \{\omega_{ij}\}}(P_i)\pi(P_i^c)}{\pi_{P_i^c\cup D'_{\mu_j}\setminus \{\omega_{ij}\}}(P_i^c)}\right).$$
	Since $(P_i\setminus (D\cup D'))\cap \{\omega_{ij}\}=\emptyset$ for each $\omega_{ij}\in D$, by Axiom \ref{axiom:sbm}, 
	\begin{multline*}
		\frac{\pi_{P_i^c\cup D'_{\mu_j}}(P_i)\pi(P_i^c)}{\pi_{P_i^c\cup D'_{\mu_j}}(P_i^c)} + \frac{\pi_{(P_i^c\cup D'_{\mu_j})\setminus ((P_i\setminus (D\cup D')) \cup \{\omega_{ij}\})}(P_i)\pi(P_i^c)}{\pi_{(P_i^c\cup D'_{\mu_j})\setminus ((P_i\setminus (D\cup D')) \cup \{\omega_{ij}\})}(P_i^c)} \leq \\
		\frac{\pi_{(P_i^c\cup D'_{\mu_j})\setminus (P_i\setminus (D\cup D'))}(P_i)\pi(P_i^c)}{\pi_{(P_i^c\cup D'_{\mu_j})\setminus (P_i\setminus (D\cup D'))}(P_i^c)}
		+ \frac{\pi_{P_i^c\cup D'_{\mu_j}\setminus \{\omega_{ij}\}}(P_i)\pi(P_i^c)}{\pi_{P_i^c\cup D'_{\mu_j}\setminus \{\omega_{ij}\}}(P_i^c)}.
	\end{multline*}
	Similarly, since $D\cap D' = \emptyset$, by Axiom \ref{axiom:sbm},
	$$\frac{\pi_{P_i^c\cup D\cup D'}(P_i)}{\pi_{P_i^c\cup D\cup D'}(P_i^c)}\leq  \frac{\pi_{P_i^c\cup D'}(P_i)}{\pi_{P_i^c\cup D'}(P_i^c)}+\frac{\pi_{P_i^c\cup D}(P_i)}{\pi_{P_i^c\cup D}(P_i^c)}.$$
	Therefore, combining the previous two expressions, 
	{\smaller
		\begin{align*}
			\pi^{\mu_{D',P_i}}(D) 
			& =	\sum_{\omega_{ij}\in D}\left(\frac{\pi_{P_i^c\cup D'_{\mu_j}}(P_i)\pi(P_i^c)}{\pi_{P_i^c\cup D'_{\mu_j}}(P_i^c)}-\frac{\pi_{P_i^c\cup D'_{\mu_j}\setminus \{\omega_{ij}\}}(P_i)\pi(P_i^c)}{\pi_{P_i^c\cup D'_{\mu_j}\setminus \{\omega_{ij}\}}(P_i^c)}\right) \\
			& \leq \sum_{\omega_{ij}\in D}\left(\frac{\pi_{(P_i^c\cup D'_{\mu_j})\setminus (P_i\setminus (D\cup D'))}(P_i)\pi(P_i^c)}{\pi_{(P_i^c\cup D'_{\mu_j})\setminus (P_i\setminus (D\cup D'))}(P_i^c)}-\frac{\pi_{(P_i^c\cup D'_{\mu_j})\setminus ((P_i\setminus (D\cup D')) \cup \{\omega_{ij}\})}(P_i)\pi(P_i^c)}{\pi_{(P_i^c\cup D'_{\mu_j})\setminus ((P_i\setminus (D\cup D')) \cup \{\omega_{ij}\})}(P_i^c)}\right) \\
			& = \frac{\pi_{P_i^c\cup D\cup D'}(P_i)\pi(P_i^c)}{\pi_{P_i^c\cup D\cup D'}(P_i^c)}- \frac{\pi_{P_i^c\cup D'}(P_i)\pi(P_i^c)}{\pi_{P_i^c\cup D'}(P_i^c)} \\
			& \leq  \frac{\pi_{P_i^c\cup D}(P_i)\pi(P_i^c)}{\pi_{P_i^c\cup D}(P_i^c)} = \pi^{D}(D).
		\end{align*}
	}
	
	\noindent\textbf{Case 4:} $D'\cap D \neq \emptyset$, $D\setminus D'\neq \emptyset$, $D'\setminus D\neq \emptyset$. By definition, 
	\begin{align*}
		\pi^{\mu_{D',P_i}}(D)
		= & \frac{\pi_{P_i^c\cup D'}(P_i)\pi(P_i^c)\pi(D\cap D')}{\pi_{P_i^c\cup D'}(P_i^c)\pi(D')} \\
		+ &  \sum_{\omega_{ij}\in D\setminus D'}\left(\frac{\pi_{P_i^c\cup D'_{\mu_j}}(P_i)\pi(P_i^c)}{\pi_{P_i^c\cup D'_{\mu_j}}(P_i^c)}-\frac{\pi_{P_i^c\cup D'_{\mu_j}\setminus \{\omega_{ij}\}}(P_i)\pi(P_i^c)}{\pi_{P_i^c\cup D'_{\mu_j}\setminus \{\omega_{ij}\}}(P_i^c)}\right).
	\end{align*}
	Similar to the previous case, Axiom \ref{axiom:sbm} implies that the second term must be smaller than
	$$ \frac{\pi_{P_i^c\cup D\cup D'}(P_i)\pi(P_i^c)}{\pi_{P_i^c\cup D\cup D'}(P_i^c)}- \frac{\pi_{P_i^c\cup D'}(P_i)\pi(P_i^c)}{\pi_{P_i^c\cup D'}(P_i^c)}.$$
	Hence, combining the previous two expressions,
	\begin{align*}
		\pi^{\mu_{D',P_i}}(D) & \leq \frac{\pi_{P_i^c\cup D'}(P_i)\pi(P_i^c)\pi(D\cap D')}{\pi_{P_i^c\cup D'}(P_i^c)\pi(D')} + \frac{\pi_{P_i^c\cup D\cup D'}(P_i)\pi(P_i^c)}{\pi_{P_i^c\cup D\cup D'}(P_i^c)}- \frac{\pi_{P_i^c\cup D'}(P_i)\pi(P_i^c)}{\pi_{P_i^c\cup D'}(P_i^c)} \\
		& = \frac{\pi_{P_i^c\cup D\cup D'}(P_i)\pi(P_i^c)}{\pi_{P_i^c\cup D\cup D'}(P_i^c)} -  \frac{\pi_{P_i^c\cup D'}(P_i)\pi(P_i^c)\pi(D'\setminus D)}{\pi_{P_i^c\cup D'}(P_i^c)\pi(D')}.
	\end{align*}
	On the other hand, by Axiom \ref{axiom:btw},
	{\small
		$$\frac{\pi_{P_i^c\cup D\cup D'}(P_i)\pi(D')}{\pi_{P_i^c\cup D\cup D'}(P_i^c) \pi(D\cup D')} \leq  \frac{\pi_{P_i^c\cup D'}(P_i)}{\pi_{P_i^c\cup D'}(P_i^c)} \:\: \Rightarrow \:\: \frac{\pi_{P_i^c\cup D\cup D'}(P_i)}{\pi_{P_i^c\cup D\cup D'}(P_i^c) \pi(D\cup D')} \leq  \frac{\pi_{P_i^c\cup D'}(P_i)}{\pi_{P_i^c\cup D'}(P_i^c)\pi(D')} ,$$
	}which implies
	$$ \frac{\pi_{P_i^c\cup D\cup D'}(P_i)\pi(P_i^c)\pi(D'\setminus D)}{\pi_{P_i^c\cup D\cup D'}(P_i^c) \pi(D\cup D')} \leq  \frac{\pi_{P_i^c\cup D'}(P_i)\pi(P_i^c)\pi(D'\setminus D)}{\pi_{P_i^c\cup D'}(P_i^c)\pi(D')}$$
	Therefore, 
	\begin{align*}
		\pi^{\mu_{D',P_i}}(D) &\leq  \frac{\pi_{P_i^c\cup D\cup D'}(P_i)\pi(P_i^c)}{\pi_{P_i^c\cup D\cup D'}(P_i^c)} -  \frac{\pi_{P_i^c\cup D'}(P_i)\pi(P_i^c)\pi(D'\setminus D)}{\pi_{P_i^c\cup D'}(P_i^c)\pi(D')} \\
		&\leq \frac{\pi_{P_i^c\cup D\cup D'}(P_i)\pi(P_i^c)}{\pi_{P_i^c\cup D\cup D'}(P_i^c)} -  \frac{\pi_{P_i^c\cup D\cup D'}(P_i)\pi(P_i^c)\pi(D'\setminus D)}{\pi_{P_i^c\cup D\cup D'}(P_i^c) \pi(D\cup D')}  \\
		&= \frac{\pi_{P_i^c\cup D\cup D'}(P_i)\pi(P_i^c)\pi(D)}{\pi_{P_i^c\cup D\cup D'}(P_i^c)\pi(D\cup D')}\\
		& \leq  \frac{\pi_{P_i^c\cup D}(P_i)\pi(P_i^c)}{\pi_{P_i^c\cup D}(P_i^c)}  = \pi^D(D),
	\end{align*}
	where the last inequality is due to Axiom \ref{axiom:btw}. Hence, in all cases, we have $\pi^{\mu_{D',P_i}}(D) \leq \pi^{D}(D)$. This concludes the proof of the claim. 
\end{proof}

Next, for any event $A\in \mathcal{A}$, we pick a prior from $\mathbb{N}$ such that $\pi_A$ is the Bayesian posterior of this prior. Let $\pi^A$ be defined by 
$$\pi^A|_{P_i} = \pi^{A\cap P_i}|_{P_i} \text{ for all } P_i\in \mathcal{P}_{\mathcal{I}} \text{ with } A\cap P_i\neq \emptyset$$
and $\pi^A|_{P_j} = \pi|_{P_j}$ for $P_j\in \mathcal{P}_{\mathcal{I}}$ with $A\cap P_j=\emptyset$. 
By construction, $\pi^A\in \mathbb{N}$. Since, by the previous claim, $\pi^{A\cap P_i}$ maximizes the likelihood of $A\cap P_i$ within $\mathbb{N}$ for each $P_i\in \mathcal{P}_{\mathcal{I}}$ with $A\cap P_i\neq \emptyset$, it follows that $\pi^A$ maximizes the likelihood of the event $A$ in $\mathbb{N}$. Subsequent claims show that $\pi_A$ is the Bayesian posterior of $\pi^A$ for each $A\in \mathcal{A}$. Note that since Axiom \ref{axiom:regularity} requires $\pi_A(A^c)=0$, to prove that $\pi_A$ is the Bayesian posterior of $\pi^A$, it suffices to show that they agree on the relative likelihoods of any two states within $A$. 

\begin{claim}
	Let $A\in \mathcal{A}$ be such that $A\subseteq P_i$ for some $P_i\in \mathcal{P}_{\mathcal{I}}$. Then, $\pi_A$ is the Bayesian posterior of $\pi^A$. 
\end{claim}

\begin{proof}
	By Axiom \ref{axiom:ideal}, $A$ is an ideal event, and hence
	$$\frac{\pi_A(\omega)}{\pi_A(\omega')}= \frac{\pi(\omega)}{\pi(\omega')}$$
	for any $\omega,\omega'\in A$. By construction, we also have 
	$$\frac{\pi(\omega)}{\pi(\omega')} = \frac{\pi^A(\omega)}{\pi^{A}(\omega')}.$$
	This shows that $\pi_A$ is the Bayesian  posterior of $\pi^A$, as desired.
\end{proof}

\begin{claim}
	Let $A\in \mathcal{A}$ be such that $A = P_i^c \cup D$ for some $P_i\in \mathcal{P}_{\mathcal{I}}$ and $D\subsetneq P_i$. Then, $\pi_A$ is the Bayesian posterior of $\pi^A.$
\end{claim}

\begin{proof}
	By construction, 
	$$\frac{\pi^A(D)}{\pi^A(P_i^c)} = \frac{\frac{\pi_A(P_i)\pi(P_i^c)}{\pi_A(P_i^c)}}{\pi(P_i^c)} = \frac{\pi_A(P_i)}{\pi_A(P_i^c)}.$$
	In addition, for any two states $\omega,\omega'\in D$, we have 
	$$\frac{\pi^A(\omega)}{\pi^A(\omega')}= \frac{\pi(\omega)}{\pi(\omega')} = \frac{\pi_A(\omega)}{\pi_A(\omega')},$$
	where the second equality is due to Axiom \ref{axiom:ideal}. 
	The same also holds for any two states $\omega,\omega' \in P_i^c$. Hence, $\pi_A$ is the Bayesian posterior of $\pi^A$, as desired. 
\end{proof}

\begin{claim}
	Suppose $\pi_A$ is the Bayesian posterior of $\pi^A$ for any event $A$ containing at least $k$ states. Then, $\pi_A$ is also the Bayesian posterior of $\pi^A$ when the number of states in $A$ is exactly $k-1$.
\end{claim}

\begin{proof}
	If $A= P_i^c\cup D$ for some $D\subseteq P_i\in \mathcal{P}_{\mathcal{I}}$ or $A\subseteq P_i$, then the result follows from the previous two claims. Hence, we can assume that if $A\cap P_i\neq \emptyset$, then $A\cap P_i^c\neq \emptyset$ and $P_i^c\setminus A\neq \emptyset$. First, for any two states $\omega,\omega'\in A\cap P_i$ for some $P_i\in \mathcal{P}_{\mathcal{I}}$, the construction and Axiom \ref{axiom:ideal} guarantee that
	$$\frac{\pi^A(\omega)}{\pi^A(\omega')}= \frac{\pi(\omega)}{\pi(\omega')} = \frac{\pi_A(\omega)}{\pi_A(\omega')}.$$
	Hence, to prove the claim, it suffices to show that
	$$\frac{\pi_A(P_i)}{\pi_A(P_i^c)}  = \frac{\pi^A(A\cap P_i)}{\pi^A(A\cap P_i^c)}=  \frac{\pi^{A\cap P_i}(A\cap P_i)}{\sum_{j\neq i: A\cap P_j\neq \emptyset} \pi^{A\cap P_j}(A\cap P_j)}$$
	for each $P_i$ with $A\cap P_i\neq \emptyset$. There are a few cases to consider.
	
	First, assume $P_i\setminus A\neq \emptyset$. Since we also have $P_i^c\setminus A\neq \emptyset$, by Axiom \ref{axiom:lmd},
	$$\frac{\pi(P_i)}{\pi(P_i^c)} \frac{\pi_A(P_i)}{\pi_A(P_i^c)} = 
	\frac{\pi_{A\cup P_i}(P_i)}{\pi_{A\cup P_i}(P_i^c)} 
	\frac{\pi_{A\cup P_i^c}(P_i)}{\pi_{A\cup P_i^c}(P_i^c)}
	\:\: \Rightarrow \:\: 
	\frac{\pi_A(P_i)}{\pi_A(P_i^c)} = 
	\frac{\pi_{A\cup P_i}(P_i)}{\pi_{A\cup P_i}(P_i^c)} 
	\frac{\pi_{A\cup P_i^c}(P_i)}{\pi_{A\cup P_i^c}(P_i^c)}
	\frac{\pi(P_i^c)}{\pi(P_i)}.$$
	Since $|A\cup P_i|\geq k$ and $|A\cup P_i^c|\geq k$, by the induction hypothesis,
	$$ \frac{\pi_{A\cup P_i}(P_i)}{\pi_{A\cup P_i}(P_i^c)} 
	\frac{\pi_{A\cup P_i^c}(P_i)}{\pi_{A\cup P_i^c}(P_i^c)}
	\frac{\pi(P_i^c)}{\pi(P_i)}
	= \frac{\pi(P_i)}{\sum_{j\neq i: A\cap P_j\neq \emptyset} \pi^{A\cap P_j}(A\cap P_j)}
	\frac{\pi^{A\cap P_i}(A\cap P_i)}{\pi(P_i^c)}
	\frac{\pi(P_i^c)}{\pi(P_i)},$$
	where I use the following expressions: $\pi^{(A\cup P_i)\cap P_i}((A\cup P_i)\cap P_i)=\pi^{P_i}(P_i)=\pi(P_i)$, $\pi^{(A\cup P_i^c)\cap P_i^c}((A\cup P_i^c)\cap P_i^c)=\pi^{P_i^c}(P_i^c)=\pi(P_i^c)$, $(A\cup P_i^c)\cap P_i = A\cap P_i$, and $(A\cup P_i)\cap P_i^c = A\cap P_i^c$. 
	Hence, using the previous two equalities, we get
	$$\frac{\pi_A(P_i)}{\pi_A(P_i^c)} = 
	\frac{\pi_{A\cup P_i}(P_i)}{\pi_{A\cup P_i}(P_i^c)} 
	\frac{\pi_{A\cup P_i^c}(P_i)}{\pi_{A\cup P_i^c}(P_i^c)}
	\frac{\pi(P_i^c)}{\pi(P_i)}
	= \frac{\pi^{A\cap P_i}(A\cap P_i)}{\sum_{j\neq i: A\cap P_j\neq \emptyset} \pi^{A\cap P_j}(A\cap P_j)},$$
	which is the desired result.
	
	Now, consider the case where $A\cap P_i=P_i$ for some $P_i\in \mathcal{P}_{\mathcal{I}}$. If $P_i$ is the only such partition element, then we already know that for any $P_j$ with $A\cap P_j\neq \emptyset$ and $P_j\setminus A\neq \emptyset$,
	$$\frac{\pi_A(P_j)}{\pi_A(P_j^c)}
	= \frac{\pi^{A\cap P_j}(A\cap P_j)}{\pi(P_i)+\sum_{l\notin \{i,j\}: A\cap P_l\neq \emptyset} \pi^{A\cap P_l}(A\cap P_l)},$$
	where I use $\pi^{A\cap P_i}(A\cap P_i) = \pi(P_i)$, as $A\cap P_i=P_i$. Since the above equality holds for all $j\neq i$ with $A\cap P_j\neq \emptyset$, we must also have 
	$$\frac{\pi_A(P_i)}{\pi_A(P_i^c)}
	= \frac{\pi(P_i)}{\sum_{j\neq i: A\cap P_j\neq \emptyset} \pi^{A\cap P_j}(A\cap P_j)}.$$
	
	In general, let $I$ be the index set such that $A\cap P_l=P_l$ for each $l\in I$. If there exists no $j\notin I$ with $A\cap P_j\neq \emptyset$, then $\cup_{l\in I} P_l \in \sigma(\mathcal{P}_{\mathcal{I}})$ is an ideal event. Hence, it must be the case that for $i\in I$,
	$$\frac{\pi_A(P_i)}{\pi_A(P_i^c)} = \frac{\pi_A(P_i)}{\pi_A(\cup_{l\in I\setminus \{i\}} P_l)} = \frac{\pi(P_i)}{\pi(\cup_{l\in I\setminus \{i\}} P_l)} =\frac{\pi^{P_i}(P_i)}{\sum_{l\in I\setminus \{i\}} \pi^{P_l}(P_l)},$$
	as desired. 
	
	Alternatively, suppose $A\cap P_j\neq \emptyset$ for some $j\notin I$. By a previous argument, for any $j\notin I$ with $A\cap P_j\neq \emptyset$,
	\begin{align*}
		\frac{\pi_A(P_j)}{\pi_A(P_j^c)}
		&= \frac{\pi^{A\cap P_j}(A\cap P_j)}{\sum_{i\in I}\pi(P_i)+\sum_{l\notin I\cup \{j\}: A\cap P_l\neq \emptyset} \pi^{A\cap P_l}(A\cap P_l)} \\
		&= \frac{\pi^{A\cap P_j}(A\cap P_j)}{\pi(\cup_{i\in I} P_i)+\sum_{ l \notin I\cup \{j\}: A\cap P_l\neq \emptyset} \pi^{A\cap P_l}(A\cap P_l)}.
	\end{align*}
	Since this is true for each $j\notin I$ with $A\cap P_j\neq \emptyset$, we must also have 
	$$\frac{\pi_A(\cup_{l\in I} P_l)}{\pi_A((\cup_{l\in I} P_l)^c)} 
	= \frac{\pi(\cup_{l\in I} P_l)}{\sum_{j\notin I: A\cap P_j\neq \emptyset} \pi^{A\cap P_j}(A\cap P_j)}.$$
	By a previous argument, idealness of $\cup_{l\in I} P_l$ implies that for $i\in I$,
	$$\frac{\pi_A(P_i)}{\pi_A(\cup_{l\in I} P_l)} =  \frac{\pi(P_i)}{\pi(\cup_{l\in I} P_l)} $$
	Combining the last two equations, for each $i\in I$, we get
	\begin{align*}
		\frac{\pi_A(P_i)}{\pi_A((\cup_{l\in I} P_l)^c)} 
		& =  \frac{\pi_A(P_i)}{\pi_A(\cup_{l\in I} P_l)}\frac{\pi_A(\cup_{l\in I} P_l)}{\pi_A((\cup_{l\in I} P_l)^c)}
		= \frac{\pi(P_i)}{\pi(\cup_{l\in I} P_l)} 
		\frac{\pi(\cup_{l\in I} P_l)}{\sum_{j\notin I: A\cap P_j\neq \emptyset} \pi^{A\cap P_j}(A\cap P_j)} \\
		& = \frac{\pi(P_i)}{\sum_{j\notin I: A\cap P_j\neq \emptyset} \pi^{A\cap P_j}(A\cap P_j)}.
	\end{align*}
	Lastly, for $i\in I$, the last two equations imply
	\begin{align*}
		\frac{\pi_A(P_i^c)}{\pi_A(P_i)} & = \frac{\pi_A(\cup_{l\in I\setminus \{i\}} P_l)}{\pi_A(P_i)} + \frac{\pi_A((\cup_{l\in I} P_l)^c)}{\pi_A(P_i)} \\
		& =  \frac{\pi(\cup_{l\in I\setminus \{i\}} P_l)}{\pi(P_i)} +  \frac{\sum_{j\notin I: A\cap P_j\neq \emptyset} \pi^{A\cap P_j}(A\cap P_j)}{\pi(P_i)}\\
		& = \frac{\sum_{j\neq i: A\cap P_j\neq \emptyset} \pi^{A\cap P_j}(A\cap P_j)}{\pi(P_i)},
	\end{align*}
	where I use $\pi^{A\cap P_l}(A\cap P_l)=\pi(P_l)$ for $l\in I$. Hence, we were able to show that 
	$$\frac{\pi_A(P_i)}{\pi_A(P_i^c)}  = \frac{\pi^A(A\cap P_i)}{\pi^A(A\cap P_i^c)}=  \frac{\pi^{A\cap P_i}(A\cap P_i)}{\sum_{j\neq i: A\cap P_j\neq \emptyset} \pi^{A\cap P_j}(A\cap P_j)}$$
	for all $P_i\in \mathcal{P}_{\mathcal{I}}$ with $A\cap P_i\neq \emptyset$. This concludes the proof of the claim. 
\end{proof}

By Claim \ref{claim:ML}, $\pi^A (A\cap P_i) = \pi^{A\cap P_i} (A\cap P_i)= \max_{\pi'\in \mathbb{N}}\pi'(A\cap P_i) = \overline{\pi}(A\cap P_i)$ for each $A\in \mathcal{A}$ and $P_i\in \mathcal{P}_{\mathcal{I}}$ with $A\cap P_i\neq \emptyset$. In addition, by construction, for any $\omega,\omega' \in A\cap P_i$, 
$$\frac{\pi^A(\omega)}{\pi^A(\omega')}= \frac{\pi(\omega)}{\pi(\omega')}.$$
Since the previous claims show that $\pi_A$ is the Bayesian posterior of $\pi^A$ for each $A\in \mathcal{A}$, this guarantees that the representation for RML updating given by equation \ref{equation:two_stage2} holds with this choice of $\mathbb{N}$. 

Next, we need to show that $\mathbb{N}$ is a consistent set of plausible priors that satisfies the richness property. 

\begin{claim}
	\label{claim:richness}
	$\mathbb{N}$ is a consistent set of plausible priors that satisfies richness.
\end{claim}
\begin{proof}
	
	Since $\pi^{\mu_{D,P_i}}(P_j) = \pi(P_j)$ for each $D\subseteq P_i\in \mathcal{P}_{\mathcal{I}}$, $\mu_{D,P_i}\in \mathcal{M}_{D,P_i}$, and $P_j\in \mathcal{P}_{\mathcal{I}}$, it follows that $\mathbb{N}\subseteq \mathbb{N}_{\mathcal{P}_{\mathcal{I}}}(\pi)$. $\mathbb{N}$ is also clearly closed and convex and $\pi\in \mathbb{N}$. Hence, $\mathbb{N}$ is a consistent set of plausible priors. Next, we need to show that it satisfies richness. 

	\begin{enumerate}
		
		\item (Non-extremeness) We want to show that there exist $\pi',\pi''\in \mathbb{N}$ satisfying $\pi'(D)<\pi(D)<\pi''(D)$ whenever $D\subsetneq P_i$ for some $P_i\in \mathcal{P}_{\mathcal{I}}$. Let $D\subsetneq P_i$ for some $P_i\in \mathcal{P}_{\mathcal{I}}$ be given. I will show that by letting $\pi' = \pi^{P_i\setminus D}$ and $\pi'' = \pi^{D}$, we get the desired result. 
		
		First, consider the case where $D = P_i\setminus \{\omega\}$ for some $\omega\in P_i$. Since $\Omega\setminus \{\omega\} \notin \sigma(\mathcal{P}_{\mathcal{I}})$ and $\{\omega\}$ is trivially ideal, it must be that $\Omega\setminus \{\omega\}$ is not ideal. This is possible only if
		$$\frac{\pi_{\Omega\setminus \{\omega\}}(P_i)}{\pi_{\Omega\setminus \{\omega\}}(P_i^c)}\neq \frac{\pi(P_i\setminus \{\omega\})}{\pi(P_i^c)}.$$
		By Axiom \ref{axiom:btw}, 
		$$\frac{\pi_{\Omega\setminus \{\omega\}}(P_i)}{\pi_{\Omega\setminus \{\omega\}}(P_i^c)}\geq  \frac{\pi(P_i\setminus \{\omega\})}{\pi(P_i^c)},$$
		so it must be that this inequality holds strictly. This shows that
		$$\pi^{P_i\setminus \{\omega\}}(P_i\setminus \{\omega\})=\frac{\pi_{\Omega\setminus \{\omega\}}(P_i)\pi(P_i^c)}{\pi_{\Omega\setminus \{\omega\}}(P_i^c)}> \pi(P_i\setminus \{\omega\}).$$
		Since $\pi^{P_i\setminus \{\omega\}}(P_i) =\pi(P_i)$, this also shows that $\pi^{P_i\setminus \{\omega\}}(\omega)<\pi(\omega)$. 
		
		Now, let $D = P_i\setminus \{\omega_1,\ldots,\omega_k\}$ with $\{\omega_1,\ldots,\omega_k\}\subsetneq P_i$ be given. By Axiom \ref{axiom:ideal}, $\{\omega_1,\ldots,\omega_k\}$ is an ideal event, so it must be that $\Omega\setminus \{\omega_1,\ldots,\omega_k\}$ is not ideal. In addition, Axiom \ref{axiom:btw} implies 
		$$\frac{\pi_{\Omega\setminus \{\omega_1,\ldots,\omega_{k}\}}(P_i)}{\pi_{\Omega\setminus \{\omega_1,\ldots,\omega_{k}\}}(P_i^c)}
		\geq 
		\frac{\pi_{\Omega\setminus \{\omega_1\}}(P_i\setminus \{\omega_1,\ldots,\omega_{k}\})}{\pi_{\Omega\setminus \{\omega_1\}}(P_i^c)}.$$
		By the previous argument, 
		$$\frac{\pi_{\Omega\setminus \{\omega_1\}}(P_i)}{\pi_{\Omega\setminus \{\omega_1\}}(P_i^c)}>  \frac{\pi(P_i\setminus \{\omega_1\})}{\pi(P_i^c)}.$$
		Then,
		\begin{align*}
			\frac{\pi_{\Omega\setminus \{\omega_1\}}(P_i\setminus \{\omega_1,\ldots,\omega_{k}\})}{\pi_{\Omega\setminus \{\omega_1\}}(P_i^c)}
			&= \frac{\pi_{\Omega\setminus \{\omega_1\}}(P_i\setminus \{\omega_1,\ldots,\omega_{k}\})}{\pi_{\Omega\setminus \{\omega_1\}}(P_i)}
			\frac{\pi_{\Omega\setminus \{\omega_1\}}(P_i)}{\pi_{\Omega\setminus \{\omega_1\}}(P_i^c)}\\
			& >\frac{\pi_{\Omega\setminus \{\omega_1\}}(P_i\setminus \{\omega_1,\ldots,\omega_{k}\})}{\pi_{\Omega\setminus \{\omega_1\}}(P_i)}
			\frac{\pi(P_i\setminus \{\omega_1\})}{\pi(P_i^c)}\\
			& = \frac{\pi(P_i\setminus \{\omega_1,\ldots,\omega_{k}\})}{\pi(P_i\setminus \{\omega_1\})}
			\frac{\pi(P_i\setminus \{\omega_1\})}{\pi(P_i^c)}\\
			& =  \frac{\pi(P_i\setminus \{\omega_1,\ldots,\omega_{k}\})}{\pi(P_i^c)},
		\end{align*}
		where the inequality is due to the previous argument and the second equality is due to the fact that $P_i\setminus \{\omega_1,\ldots,\omega_k\}$ is an ideal event by Axiom \ref{axiom:ideal}. This shows that
		$$\frac{\pi_{\Omega\setminus \{\omega_1,\ldots,\omega_k\}}(P_i)}{\pi_{\Omega\setminus \{\omega_1,\ldots,\omega_k\}}(P_i^c)}> \frac{\pi(P_i\setminus \{\omega_1,\ldots,\omega_k\})}{\pi(P_i^c)}.$$
		Lastly, by construction, 
		$$\pi^{P_i\setminus \{\omega_1,\ldots,\omega_k\}}(P_i\setminus \{\omega_1,\ldots,\omega_k\})
		= \frac{\pi_{\Omega\setminus \{\omega_1,\ldots,\omega_k\}}(P_i) \pi(P_i^c)}{\pi_{\Omega\setminus \{\omega_1,\ldots,\omega_k\}}(P_i^c)}
		> \pi(P_i\setminus \{\omega_1,\ldots,\omega_k\}).$$
		This also implies that $\pi^{P_i\setminus \{\omega_1,\ldots,\omega_k\}}(\{\omega_1,\ldots,\omega_k\})<\pi(\{\omega_1,\ldots,\omega_k\})$, as $ \pi(P_i) = \pi^{P_i\setminus \{\omega_1,\ldots,\omega_k\}}(P_i)$. Now, we have shown $\pi^D(D)>\pi(D)$ and $\pi^{P_i\setminus D}(D)<\pi(D)$, which is the desired result.
		
		\item (Prior-coherence) By Claim \ref{claim:ML}, $\pi^A$ maximizes the likelihood of the event $A$ within $\mathbb{N}$, and, by construction, it satisfies 
		$$ \frac{\pi^A(\omega)}{\pi^{A}(\omega')} = \frac{\pi(\omega)}{\pi(\omega')}$$
		for each $\omega,\omega'\in A\cap P_i$ with $A\cap P_i\neq \emptyset$. 
		
		\item (Chain-coherence) Let $P_i\in \mathcal{P}_{\mathcal{I}}$ and enumerate the states in $P_i$ as $\{\omega_{i1},\ldots, \omega_{ik_i}\}$. Let $\mu:\{1,\ldots,k_i\} \rightarrow \{1,\ldots,k_i\}$ denote a permutation function, and let $\mathcal{M}$ denote the collection of all such permutations. Clearly, chain-coherence holds if and only if it holds for chains of the type $\{\omega_{i\mu(1)}\} \subseteq \{\omega_{i\mu(1)}, \omega_{i\mu(2)}\} \subseteq \cdots \subseteq \{\omega_{i\mu(1)}, \ldots, \omega_{i\mu(k_i)}\}$ for any permutation $\mu \in \mathcal{M}$. Let $D=\{\omega_{i\mu(1)}\}$ and $\mu_{D,P_i}= \mu^{-1}$. We need to show that $\pi^{\mu_{D,P_i}}$ maximizes the likelihood of each event in the chain. First, we already know that $\pi^{\mu_{D,P_i}}$ maximizes the likelihood of $\{\omega_{i\mu(1)}\}$ by Claim \ref{claim:ML}. Consider the event $D'=\{\omega_{i\mu(1)}, \ldots,\omega_{i\mu(m)} \}$ where $m\in \{2,\ldots,k_i\}$. By construction,
		{\small 
			\begin{align*}
				\pi^{\mu_{D,P_i}}(D') 
				= &\frac{\pi_{P_i^c\cup \{\omega_{i\mu(1)}\}}(P_i)\,\pi(P_i^c)}{\pi_{P_i^c\cup \{\omega_{i\mu(1)}\}}(P_i^c)} \\
				+ & \sum_{j=1}^{m-1}\left(
				\frac{\pi_{P_i^c\cup \{\omega_{i\mu(1)}, \dots,\omega_{i\mu(j+1)}\}}(P_i)\,\pi(P_i^c)}
				{\pi_{P_i^c\cup \{\omega_{i\mu(1)}, \dots,\omega_{i\mu(j+1)}\}}(P_i^c)} - \frac{\pi_{P_i^c\cup \{\omega_{i\mu(1)}, \dots,\omega_{i\mu(j)}\}}(P_i)\,\pi(P_i^c)}
				{\pi_{P_i^c\cup \{\omega_{i\mu(1)}, \dots,\omega_{i\mu(j)}\}}(P_i^c)}
				\right)\\
				= & \frac{\pi_{P_i^c\cup \{\omega_{i\mu(1)}, \dots,\omega_{i\mu(m)}\}}(P_i)\,\pi(P_i^c)}
				{\pi_{P_i^c\cup \{\omega_{i\mu(1)}, \dots,\omega_{i\mu(m)}\}}(P_i^c)}\\
				= & \frac{\pi_{P_i^c\cup D'}(P_i)\pi(P_i^c)}{\pi_{P_i^c\cup D'}(P_i^c)} = \pi^{D'}(D').
			\end{align*}
		}Since $\pi^{D'}$ maximizes the likelihood of $D'$, so does $\pi^{\mu_{D,P_i}}$. This shows that $\mathbb{N}$ satisfies chain-coherence.

		\item ($\mathcal{P}$-Rectangularity) Suppose $\pi',\pi''\in \mathbb{N}$ and let $P_i\in \mathcal{P}_{\mathcal{I}}$. Then, by construction, $\pi'''$ and $\pi''''$ defined by the equations 
		$$ \pi''' |_{P_i} = \pi' |_{P_i}, \quad \pi''' |_{P_i^c} = \pi'' |_{P_i^c}, \quad  \pi'''' |_{P_i} =  \pi'' |_{P_i}, \quad \pi'''' |_{P_i^c} =  \pi' |_{P_i^c}$$
		belong to $\mathbb{N}$, as desired.
	\end{enumerate}
\end{proof}

To conclude the proof of the theorem, we need to show that if $\mathbb{N}^*$ is a rich set of plausible priors that induces $\{\pi_A\}_{A\in \mathcal{A}}$, then $\mathbb{N}^*= \mathbb{N}$. 

\begin{claim}
	\label{claim:uniqueness}
	Suppose $\{\pi_A\}_{A\in \mathcal{A}}$ is consistent with RML updating with a rich set of plausible priors $\mathbb{N}^*$.  Then, as long as $\mathcal{P}\neq \{\Omega\}$, we have $\mathbb{N}^*= \mathbb{N}$, where $\mathbb{N}$ is defined as in equation \ref{equation:priors}. 
\end{claim}

\begin{proof}
	Assume $\mathcal{P}\neq \{\Omega\}$, so that $\mathcal{P}$ is uniquely revealed as $\mathcal{P}_{\mathcal{I}}$ by Proposition \ref{proposition:ideal}. Let $\pi'\in \mathbb{N}^*$ be given and suppose $\pi'\notin \mathbb{N}$. By definition, it must be that there exists $P_i\in \mathcal{P}_{\mathcal{I}}$ such that 
	$$\pi'|_{P_i} \notin \text{co}\left (\{\pi^{\mu_{D,P_i}}|_{P_i}\}_{D\subseteq P_i \text{ and } \mu_{D,P_i}\in \mathcal{M}_{D,P_i}}\right).$$
	By the separating hyperplane theorem, we can find a vector $c = (c_1,\ldots, c_{k_i}) \in \mathbb{R}^{k_i}$ such that 
	\begin{equation}
		\label{equation:hyperplane}
		c \cdot \pi'|_{P_i} > \max_{D\subseteq P_i \text{ and } \mu_{D,P_i}\in \mathcal{M}_{D,P_i}} c \cdot \pi^{\mu_{D,P_i}}|_{P_i}.
	\end{equation}
	Let $\mu: \{1,\ldots, k_i\} \rightarrow  \{1,\ldots, k_i\}$ be a permutation that ensures $c_{\mu(j)}\geq  c_{\mu(j+1)}$ for $j\in \{1,\ldots,k_i-1\}$. Consider the chain $\{\omega_{i\mu(1)}\} \subseteq \{\omega_{i\mu(1)}, \omega_{i\mu(2)}\} \subseteq \cdots \subseteq \{\omega_{i\mu(1)}, \ldots, \omega_{i\mu(k_i)}\}$, and let $S_{\mu_j} = \{\omega_{i\mu(1)}, \ldots,\omega_{i\mu(j)}\}$. Let $D=\{\omega_{i\mu(1)}\}$ and $\mu_{D,P_i}=\mu^{-1}$. By construction, $\pi^{\mu_{D,P_i}}$ maximizes the likelihood of $S_{\mu_j}$ within $\mathbb{N}$ for all $j\in \{1,\ldots,k_i\}$, as shown in Claim \ref{claim:richness}.  In addition, since for each $S_{\mu_j}$ the maximal possible likelihood is uniquely revealed from $\{\pi_A\}_{A\in \mathcal{A}}$ and must coincide with $\pi^{\mu_{D,P_i}}(S_{\mu_j})$, we must have 
	\begin{equation}
		\label{equation:chain_bound}
		\pi'(S_{\mu_j}) \leq  \pi^{\mu_{D,P_i}} (S_{\mu_j})
	\end{equation}
	for all $j\in \{1,\ldots, k_i\}$ with equality when $j=k_i$, as all priors must agree on the unambiguous event $P_i=S_{\mu_{k_i}}$. Now notice that 
	$$c \cdot \pi'|_{P_i}  = \sum_{j=1}^{k_i} c_{\mu(j)} \pi'(\omega_{i\mu(j)}) = c_{\mu(k_i)}\pi'(S_{\mu_{k_i}}) + \sum_{j=1}^{k_i-1}\pi'(S_{\mu_j})\left(c_{\mu(j)}-c_{\mu(j+1)}\right).$$
	Similarly,
	{\small
		$$c \cdot \pi^{\mu_{D,P_i}}|_{P_i}  = \sum_{j=1}^{k_i} c_{\mu(j)} \pi^{\mu_{D,P_i}}(\omega_{i\mu(j)}) = c_{\mu(k_i)}\pi^{\mu_{D,P_i}}(S_{\mu_{k_i}}) + \sum_{j=1}^{k_i-1}\pi^{\mu_{D,P_i}}(S_{\mu_j})\left(c_{\mu(j)}-c_{\mu(j+1)}\right).$$
	}Subtracting the first term from the second one and using the fact that $\pi'(S_{\mu_{k_i}}) = \pi^{\mu_{D,P_i}}(S_{\mu_{k_i}})$, we get
	$$c \cdot \pi^{\mu_{D,P_i}}|_{P_i} -c \cdot \pi'|_{P_i}  =  \sum_{j=1}^{k_i-1} \left(\pi^{\mu_{D,P_i}}(S_{\mu_j})-\pi'(S_{\mu_j})\right)\left(c_{\mu(j)}-c_{\mu(j+1)}\right).$$
	By equation \ref{equation:chain_bound}, the term inside the first parentheses is non-negative. By the choice of $\mu$, the term inside the second parentheses is non-negative. Hence, the above term must be non-negative, which contradicts equation \ref{equation:hyperplane}. Therefore, we must have $\pi'\in \mathbb{N}$. This shows that $\mathbb{N}^*\subseteq \mathbb{N}$.
	
	To show that $\mathbb{N}\subseteq \mathbb{N}^*$, first, consider $\pi^{\mu_{D,P_i}}\in \mathbb{N}$ for some $D\subseteq P_i\in \mathcal{P}_{\mathcal{I}}$ and $\mu_{D,P_i}\in \mathcal{M}_{D,P_i}$ with $|D|=1$. To simplify the notation, if $D=\{\omega_{ij}\}$ for some $\omega_{ij}\in P_i$, let $\mu_{ij} = \mu_{D,P_i}$, $\pi^{\mu_{ij}} = \pi^{\mu_{D,P_i}}$, and $\mathcal{M}_{ij} = \mathcal{M}_{D,P_i}$. By letting $\mu = \mu_{ij}^{-1}$, we have that $\pi^{\mu_{ij}}$ maximizes the likelihood of every event in the chain 
	$\{\omega_{i\mu(1)}\} \subseteq \{\omega_{i\mu(1)}, \omega_{i\mu(2)}\} \subseteq \cdots \subseteq \{\omega_{i\mu(1)}, \ldots, \omega_{i\mu(k_i)}\}$, as was shown in Claim \ref{claim:richness}. Since the maximal likelihood of every event within the chain is uniquely determined given $\{\pi_A\}_{A\in \mathcal{A}}$ and, by chain-coherence, there must be some prior within $\mathbb{N}^*$ maximizing the likelihood of every event for this chain, it must be that $\pi^{\mu_{ij}}|_{P_i} = \pi'|_{P_i}$ for some $\pi'\in \mathbb{N}^*$.  Since $\pi^{\mu_{ij}}|_{P_i^c} = \pi|_{P_i^c}$ and $\pi\in \mathbb{N}^*$, by $\mathcal{P}$-Rectangularity of $\mathbb{N}^*$, $\pi^{\mu_{ij}}\in \mathbb{N}^*$. 
	
	Next, consider $\pi^{\mu_{D,P_i}}\in \mathbb{N}$ for some $D\subseteq P_i\in \mathcal{P}_{\mathcal{I}}$ and $\mu_{D,P_i}\in \mathcal{M}_{D,P_i}$ such that $|D|\geq 2$. By letting $\mu = \mu_{D,P_i}^{-1}$, we have that $D = \{\omega_{i\mu(1)}, \ldots, \omega_{i\mu(m)}\}$ for some $m\leq k_i$. As before, let  $S_{\mu_l} = \{\omega_{i\mu(1)}, \ldots, \omega_{i\mu(l)}\}$, and hence $D = S_{\mu_m}$. Next, let $\mathbb{N}(\mu_{D,P_i})$ denote the collection of all $\pi^{\mu_{ij}}$ such that $\mu_{ij}$ agrees with $\mu_{D,P_i}$ on $P_i\setminus D$. Now, as was shown in Claim \ref{claim:richness}, any $\pi^{\mu_{ij}}\in \mathbb{N}(\mu_{D,P_i})$ maximizes the likelihood of $S_{\mu_l}$ for $l\in \{m,\ldots,k_i\}$, which must coincide with $\pi^{\mu_{D,P_i}}(S_{\mu_l})$. This is possible only if $\pi^{\mu_{ij}}(\omega) = \pi^{\mu_{D,P_i}}(\omega)$ for all $\omega\in P_i\setminus D$. In addition, $\pi^{\mu_{D,P_i}}(D) = \pi^{\mu_{ij}}(D)$ must hold, as both maximize the likelihood of $D$. We show that 
	$$\pi^{\mu_{D,P_i}}|_{D} \in \text{co}\left(\{\pi^{\mu_{ij}}|_{D}\}_{\pi^{\mu_{ij}}\in \mathbb{N}(\mu_{D,P_i})}\right).$$
	Suppose this is not the case. Then, by the separating hyperplane theorem, we can find a vector $c=(c_1,\ldots,c_m)\in \mathbb{R}^m$ such that 
	\begin{equation}
		\label{equation:hyperplane2}
		c \cdot \pi^{\mu_{D,P_i}}|_{D} > \max_{\pi^{\mu_{ij}} \in \mathbb{N}(\mu_{D,P_i})} c \cdot \pi^{\mu_{ij}}|_{D}.
	\end{equation}
	Let $\tilde{\mu}:\{1,2,\ldots,m\}\rightarrow \{1,2,\ldots,m\}$ denote a permutation such that $c_{\tilde{\mu}(l)}\geq  c_{\tilde{\mu}(l+1)}$ for $l\in \{1,\ldots,m-1\}$. Pick a permutation $\mu_{ij}$ that induces the chain $\{\omega_{i\tilde{\mu}(1)}\} \subseteq \{\omega_{i\tilde{\mu}(1)}, \omega_{i\tilde{\mu}(2)}\} \subseteq \cdots \subseteq \{\omega_{i\tilde{\mu}(1)}, \ldots, \omega_{i\tilde{\mu}(m)}\}$, and let $S_{\tilde{\mu}_l} = \{\omega_{i\tilde{\mu}(1)}, \ldots, \omega_{i\tilde{\mu}(l)}\}$. Since $\pi^{\mu_{ij}}$ maximizes the likelihood of every event within this chain, we have that 
	\begin{equation}
		\label{equation:chain_bound2}
		\pi^{\mu_{D,P_i}}(S_{\tilde{\mu}_l})\leq \pi^{\mu_{ij}}(S_{\tilde{\mu}_l})
	\end{equation}
	for all $l\in \{1,\ldots, m\}$ with equality when $l=m$, as $S_{\tilde{\mu}_m}=D$ and $\pi^{\mu_{D,P_i}}(D) = \pi^{\mu_{ij}}(D)$. Now notice that 
	$$c \cdot \pi^{\mu_{D,P_i}}|_{D}  = \sum_{l=1}^{m} c_{\tilde{\mu}(l)} \pi^{\mu_{D,P_i}}(\omega_{i\tilde{\mu}(l)}) = c_{\tilde{\mu}(m)}\pi^{\mu_{D,P_i}}(S_{\tilde{\mu}_m}) + \sum_{l=1}^{m-1}\pi^{\mu_{D,P_i}}(S_{\tilde{\mu}_l})\left(c_{\tilde{\mu}(l)}-c_{\tilde{\mu}(l+1)}\right).$$
	Similarly,
	$$c \cdot \pi^{\mu_{ij}}|_{D} = \sum_{l=1}^{m} c_{\tilde{\mu}(l)} \pi^{\mu_{ij}}(\omega_{i\tilde{\mu}(l)}) = c_{\tilde{\mu}(m)} \pi^{\mu_{ij}}(S_{\tilde{\mu}_m}) + \sum_{l=1}^{m-1} \pi^{\mu_{ij}}(S_{\tilde{\mu}_l}) \left(c_{\tilde{\mu}(l)} - c_{\tilde{\mu}(l+1)}\right).$$
	Since $S_{\tilde{\mu}_{m}} = D$, subtracting the first term from the second one and using the fact that $\pi^{\mu_{D,P_i}}(D) = \pi^{\mu_{ij}}(D)$, we get
	$$c \cdot \pi^{\mu_{ij}}|_{D}-c \cdot \pi^{\mu_{D,P_i}}|_{D}  = \sum_{l=1}^{m-1}\left( \pi^{\mu_{ij}}(S_{\tilde{\mu}_l}) - \pi^{\mu_{D,P_i}}(S_{\tilde{\mu}_l})\right) \left(c_{\tilde{\mu}(l)} - c_{\tilde{\mu}(l+1)}\right).$$
	By equation \ref{equation:chain_bound2}, the term inside the first parentheses is non-negative. By the choice of $\tilde{\mu}$, the term inside the second parentheses is non-negative. Hence, the above term must be non-negative, which contradicts equation \ref{equation:hyperplane2}. Therefore, we must have 
	$$\pi^{\mu_{D,P_i}}|_{D} \in \text{co}\left(\{\pi^{\mu_{ij}}|_{D}\}_{\pi^{\mu_{ij}}\in \mathbb{N}(\mu_{D,P_i})}\right).$$
	Since $\pi^{\mu_{D,P_i}}$ and $\pi^{\mu_{ij}}$ agree on $P_i\setminus D$ whenever $\pi^{\mu_{ij}}\in \mathbb{N}(\mu_{D,P_i})$, and they both agree with $\pi$ on $P_i^c$, this also shows that 
	$$\pi^{\mu_{D,P_i}}\in \text{co}\left(\{\pi^{\mu_{ij}}\}_{\pi^{\mu_{ij}}\in \mathbb{N}(\mu_{D,P_i})}\right).$$
	We already have that $\pi^{\mu_{ij}}\in \mathbb{N}^*$, which is a convex set. Hence, we must have  $\pi^{\mu_{D,P_i}}\in \mathbb{N}^*$. To complete the proof of the claim, it suffices to note that if $\pi'\in \mathbb{N}$, then, by definition, $\pi'|_{P_i}$ is in the convex hull of priors $\pi^{\mu_{D,P_i}}|_{P_i}$ for each $P_i\in \mathcal{P}_{\mathcal{I}}$, so it has to be the case that $\pi'\in \mathbb{N}^*$. 
\end{proof}

\newpage 

\bibliographystyle{abbrvnat}
\bibliography{updating}

\end{document}